\documentstyle[amssymb,epsfig]{mn2e}

\newcommand{\dts}{cluster disruption time-scale}

\title[The LMC Star Cluster System] {The poorly constrained cluster
disruption time-scale in the Large Magellanic Cloud}
\author[G.~Parmentier \& R. de Grijs]{Genevi\`eve
Parmentier$^{1,2}$\thanks{E-mail: gparm@astro.uni-bonn.de}
                  \thanks{Humboldt Research Fellow} 
                  \thanks{Research Fellow of Belgian Science Policy}
                  \thanks{Honorary Scientific Research Worker of 
{\sl Fonds National de la Recherche Scientifique}, Belgium}
and Richard de Grijs$^{3,4}$ \\ 
$^1$Argelander Institute for Astronomy, University of
Bonn, Auf dem Huegel, D-53121 Bonn, Germany\\
$^2$Institute of Astrophysics \& Geophysics, University of Li\`ege, 
17 avenue du 6 Ao\^ut, B-4000 Li\`ege, Belgium \\
$^3$Department of Physics \& Astronomy, The University of Sheffield,
Hicks Building, Hounsfield Road, Sheffield S3 7RH, UK \\
$^4$National Astronomical Observatories, Chinese Academy of Sciences,
20A Datun Road, Chaoyang District, Beijing 100012, P. R. China}

\begin{document}

\date{Accepted 2007 October 16. Received 2007 October 05; in original form 2007 August 16}

\pagerange{\pageref{firstpage}--\pageref{lastpage}} \pubyear{2007}

\maketitle

\label{firstpage}
 
\begin{abstract}
We use Monte-Carlo simulations, combined with homogeneously determined
age and mass distributions based on multi-wavelength photometry, to
constrain the cluster formation history and the rate of bound cluster
disruption in the Large Magellanic Cloud (LMC) star cluster system. We
evolve synthetic star cluster systems formed with a power-law initial 
cluster mass function (ICMF) of spectral index $\alpha =-2$ assuming 
different {\dts}s.  For each of these cluster disruption
time-scales we derive the corresponding cluster formation rate (CFR) 
required to reproduce the observed cluster age distribution. 
We then compare, in a ``Poissonian'' $\chi^2$ sense, model  
mass distributions and model two-dimensional distributions in
log(mass) vs. log(age) space of the detected surviving clusters to the
observations.  Because of the bright detection limit ($M_V^{\rm lim} \simeq -4.7$ mag) 
above which the observed cluster sample is complete, one cannot constrain the
characteristic cluster disruption time-scale for a $10^4$ M$_\odot$
cluster, $t_4^{\rm dis}$ (where the disruption time-scale depends on
cluster mass as $t_{\rm dis} = t_4^{\rm dis} (M_{\rm cl} / 10^4 {\rm
M}_\odot )^\gamma$, with $\gamma \simeq 0.62$), to better than a lower
limit, $t_4^{\rm dis} \ge 1$\,Gyr. \\
We conclude that the CFR has been increasing steadily from 0.3 clusters 
Myr$^{-1}$ 5 Gyr ago, to a present rate of $(20-30)$ clusters Myr$^{-1}$,
for clusters spanning a mass range of
$\sim 100-10^7$ M$_\odot$. For older ages the derived CFR depends
sensitively on our assumption of the underlying CMF shape. If we
assume a universal Gaussian ICMF, then the CFR has increased steadily
over a Hubble time from $\sim 1$ cluster Gyr$^{-1}$ 15 Gyr ago to its
present value. On the other hand, if the ICMF has always been a
power law with a slope close to $\alpha=-2$, the CFR exhibits a
minimum some 5 Gyr ago, which we tentatively identify with the
well-known age gap in the LMC's cluster age distribution.
\end{abstract}


\section{Introduction}
\label{sec:intro}

The mass and age distributions of star cluster systems contain the
(fossil) records of their formation conditions. They are therefore
among the best tracers of the star-formation histories of their host
galaxies available to observers. It is important to realise, however,
that one needs to understand both the dominant internal and external
evolutionary processes in order to disentangle this formation record,
and hence obtain a glimpse of the initial conditions required for star
cluster formation.

The effects of stellar evolution in a given star cluster (which can be
approximated by a ``simple'' stellar population once the cluster has
reached an age that is well in excess of its formation time-scale) are
rather well understood, whereas we have only recently begun to make
major {\it quantitative} inroads into understanding the environmental
effects leading to star cluster ``weight loss'' (i.e., the
preferential depletion of the low-mass component of the cluster's
stellar mass function caused by tidal stripping and the ejection of
stars owing to internal two-body relaxation) and -- eventually --
disruption.

Estimates of the characteristic cluster disruption time-scales in
various star cluster environments have been calculated by Boutloukos
\& Lamers (2003), de Grijs et al. (2003a,b,c), Gieles et al. (2005)
and de Grijs \& Anders (2006), among others (see also Lamers et
al. 2005a,b). Specifically, Boutloukos \& Lamers (2003) and Lamers et
al. (2005b) show that the cluster age distribution and cluster mass
function approximate a double power law when a cluster system is
affected by both fading and secular dynamical evolution. Knowledge of
the detection limit in terms of cluster mass vs. age, combined with
the estimate of the cluster age or mass at the ``break point'' of the
integrated cluster age (mass) distribution (see, e.g., fig. 1 in
Boutloukos \& Lamers 2003, where the ``break points'' are referred to
as $t_{\rm cross}$ and $M_{\rm cross}$ in the age and mass
distributions, respectively) then leads to the typical cluster
disruption time-scale. Their analysis, however, explicitly builds on
the assumption of a constant cluster formation rate (CFR, i.e. the
number of clusters formed per linear time interval ${\rm d}N/{\rm d}t$
is constant in time) as a function of time. In that context, the
(poorly-known) time-variable CFR of the LMC cluster system hampers
such an analysis. If only the observed cluster age distribution is
known, then we are left with a degeneracy between the CFR and the
disruption time-scale, in the sense that one cannot distinguish
between a low CFR combined with slow secular dynamical evolution on
the one hand, and a vigourous CFR combined with cluster disruption
occurring on a rapid time-scale on the other.

The star cluster system in the Large Magellanic Cloud (LMC) has the
potential of providing strong constraints to the theory of star
cluster disruption as a function of environment, since it is composed
of the largest resolved cluster system spanning {\it both} a
reasonable mass range ($\sim 10^2 - 10^6$ M$_\odot$; cf. Hunter et
al. 2003, hereafter H03; de Grijs \& Anders 2006, and references
therein) {\it and} an age range from a few Myr to $\sim 13$ Gyr
available. In addition, thanks to the LMC's proximity, we have been
able to obtain observations -- and derived the age and mass
distributions -- of a sufficiently large cluster sample to allow a
statistical approach to its evolution (e.g., H03; de Grijs \& Anders
2006; see also Sect. 2).

On the basis of the few star clusters systems analysed in detail to
date, including M51 and the Antennae interacting system, Bastian et
al. (2005), Fall et al. (2005) and Fall (2006) suggest that the early
evolution of star cluster systems is most likely characterised by a
rapid, largely mass-independent ``infant mortality'' phase, at least
for masses $\ga 10^4$ M$_\odot$ (see also de Grijs \& Parmentier 2007;
and references therein), combined with ``infant weight loss''(the loss 
of stars caused by rapid, early gas expulsion; cf. Weidner et al. 2007),
the effects of which are enhanced by stellar evolutionary mass loss.
In this scenario, this early phase, which ends when clusters reach an 
age of $\sim 40$ to 50 Myr (e.g., Goodwin \& Bastian 2006), 
would then be followed by (mass-dependent) secular evolution. The
early, rapid cluster disruption process results from the expulsion of
the intracluster gas due to adiabatic or explosive expansion driven by
stellar winds or supernova activity (Mengel et
al. 2005; Bastian \& Goodwin 2006; Goodwin \& Bastian 2006; see de
Grijs \& Parmentier 2007 for a review). Star clusters are expected to
settle back into virial equilibrium $\sim 40$ to 50 Myr after gas
expulsion (Goodwin \& Bastian 2006). In our analysis in this paper we
will therefore exclude star clusters younger than 50 Myr, since our
main purpose is to derive the characteristic (mass-dependent)
time-scale of cluster disruption in the LMC driven by secular
evolution only. In a follow-up paper (Goodwin et al., in prep.), we
will discuss the evolution of the LMC cluster system on the shortest
time-scales relevant to the infant mortality and infant weight loss
scenarios.

In de Grijs \& Anders (2006), we found that the LMC's CFR has been
roughly constant outside of the well-known age gap between $\sim 3$
and 13 Gyr, when the CFR was a factor of $\sim 5$ lower (assuming a
roughly constant rate during this entire period). Based on this
observation as our main underlying assumption, we used a simple
approach to derive the characteristic cluster disruption time-scale in
the LMC, for which we found that $\log(t_4^{\rm dis} {\rm yr}^{-1}) =
9.9 \pm 0.1$, where $t_{\rm dis} = t_4^{\rm dis} (M_{\rm cl}/10^4 {\rm
M}_\odot)^{0.62}$ (Boutloukos \& Lamers 2003; Baumgardt \& Makino
2003; Gieles et al.~2005). We argued that this was consistent with
earlier, preliminary work done on a smaller cluster sample: Boutloukos
\& Lamers (2002) found $\log( t_4^{\rm dis} {\rm yr}^{-1} ) = 9.7 \pm
0.3$ for a smaller sample of 478 clusters within 5 kpc from the centre
of the LMC, in the age range $7.8 \le \log(\mbox{age yr}^{-1}) \le
10.0$. We also considered our result qualitatively consistent with
Hunter et al. (2003), who noticed very little destruction of clusters
at the high-mass end. This long characteristic disruption time-scale
would imply that hardly any of our LMC sample clusters are affected by
significant disruptive processes, so that we are in fact observing the
{\it initial} cluster mass function (CMF).

However, a close inspection of fig. 6 of de Grijs \& Anders (2006)
highlights an apparent contradiction. The ``crossing time'', $t_{\rm
cross}$, defined by the crossing point between the best-fitting lines
describing the number of clusters per unit time-scale that are mostly
affected by fading of their stellar populations and those that are
undergoing significant secular disruption, seems to imply that a more
appropriate time-scale for the disruption of the LMC cluster system
may be of order $\log( t_4^{\rm dis} {\rm yr}^{-1} ) \simeq
8.9$. Since this implies a downward adjustment of the characteristic
cluster disruption time-scale in the LMC by up to an order of
magnitude, we decided to re-investigate the LMC's cluster disruption
history.

Here, we approach this problem from a different angle, by running a
large number of Monte-Carlo simulations in which we vary the cluster
disruption time-scale. Meanwhile, for each of these cluster disruption
time-scales we derive the corresponding CFR required to reproduce the
observed cluster age distribution. We then match the observed cluster
mass distribution, integrated over time, and the observed
two-dimensional distribution of the detected surviving clusters in the
log(mass) vs. log(age) plane to the model results. $\chi^2$ fit
estimates are used to quantify which cluster disruption time-scale
and, therefore, which cluster formation history, best describes the
presently available data.

This paper is organised as follows. In Section \ref{sec:data} we
justify our choices used in the data analysis leading to the cluster
age and mass distributions used in the remainder of the paper. Section
\ref{sec:synth} discusses our basic assumptions in constructing
synthetic cluster populations, which we then use in Section
\ref{sec:disr} to explore the range of characteristic cluster
disruption time-scales allowed by the data. In Section \ref{sec:comp}
we highlight the importance of properly understanding the data's
completeness limit, and use this in Section \ref{sec:agap} to
constrain possible variations in the CFR over time.  In Section
\ref{sec:1vs10}, we assess precisely what we would need
to fully and unambiguously constrain $t_4^{\rm dis}$.  
Our results and conclusions are summarized in Section \ref{sec:conc}.

\begin{figure*}
\begin{minipage}[t]{\linewidth}
\centering\epsfig{figure=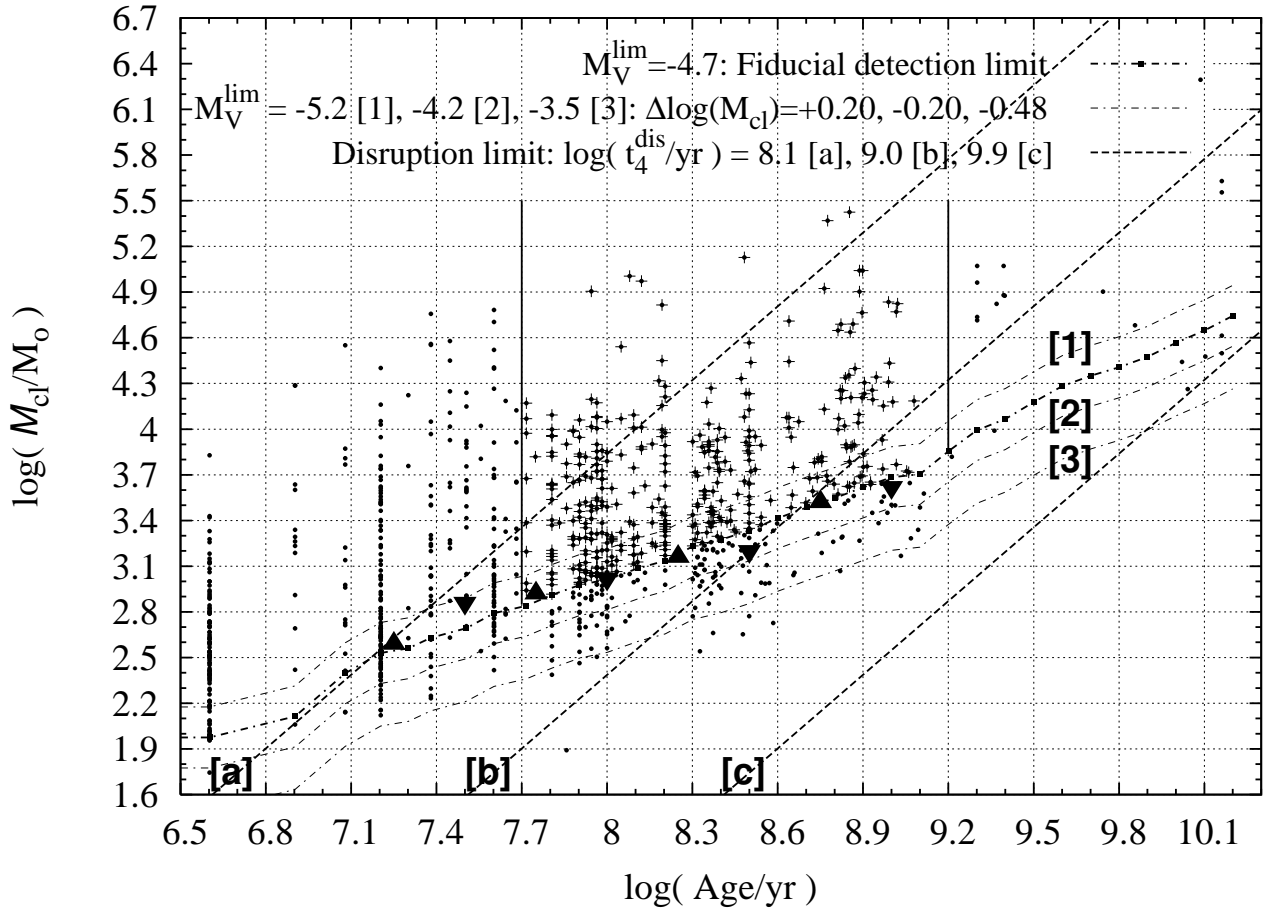, width=\linewidth}
\end{minipage}
\caption{Distribution of the LMC star cluster sample of de Grijs \&
Anders (2006) in the [$\log({\rm age}),\log(M_{\rm cl})$] plane. The
filled triangles correspond to the vertical dashed lines in the
individual panels of Fig.~\ref{fig:compl} (upright triangles:
left-hand column; upside-down triangles: right-hand column). For
subsequent cluster age ranges (in steps of 0.25 dex and 0.5 dex wide)
they trace the mass limit below which the sample becomes incomplete
(see section \ref{sec:synth} for details). They are therefore
considered tracers of the fiducial detection limit (thick
dash-dotted line with squares), which corresponds to $M_V^{\rm lim}=-4.7$ mag
(based on the {\sc galev} mass-to-light ratios for ``simple'' stellar
populations). The three thin dash-dotted lines, labelled `[1]', `[2]'
and `[3]', are the detection limits corresponding to $M_V^{\rm
lim}=-5.2$, $M_V^{\rm lim}=-4.2$ and $M_V^{\rm lim}=-3.5$ mag,
respectively.  They are therefore equivalent to the thick dash-dotted
line shifted vertically by, respectively, $\Delta \log (M_{\rm
cl})=0.2, -0.2$, and $-0.48$. The lower dash-dotted curve (`[3]') is
the $M_V^{\rm lim}=-3.5$\,mag fading limit of H03. The thick dashed
lines represent the cluster disruption limits for $\log (t_4^{\rm dis}
{\rm yr}^{-1})=8.1, 9.0$ and $9.9$ (labelled `[a]', `[b]' and `[c]',
respectively). The age range on which we focus in this paper is
bracketed by vertical solid lines, at $\log (\mbox{age yr}^{-1})=7.7$
and $\log (\mbox{age yr}^{-1})=9.2$; clusters brighter than $M_V^{\rm lim}=-4.7$ 
mag in that age range are represented by crosses}
\label{fig:obs_am} 
\end{figure*}

\begin{figure*}
\begin{minipage}[t]{\linewidth}
\centering\epsfig{figure=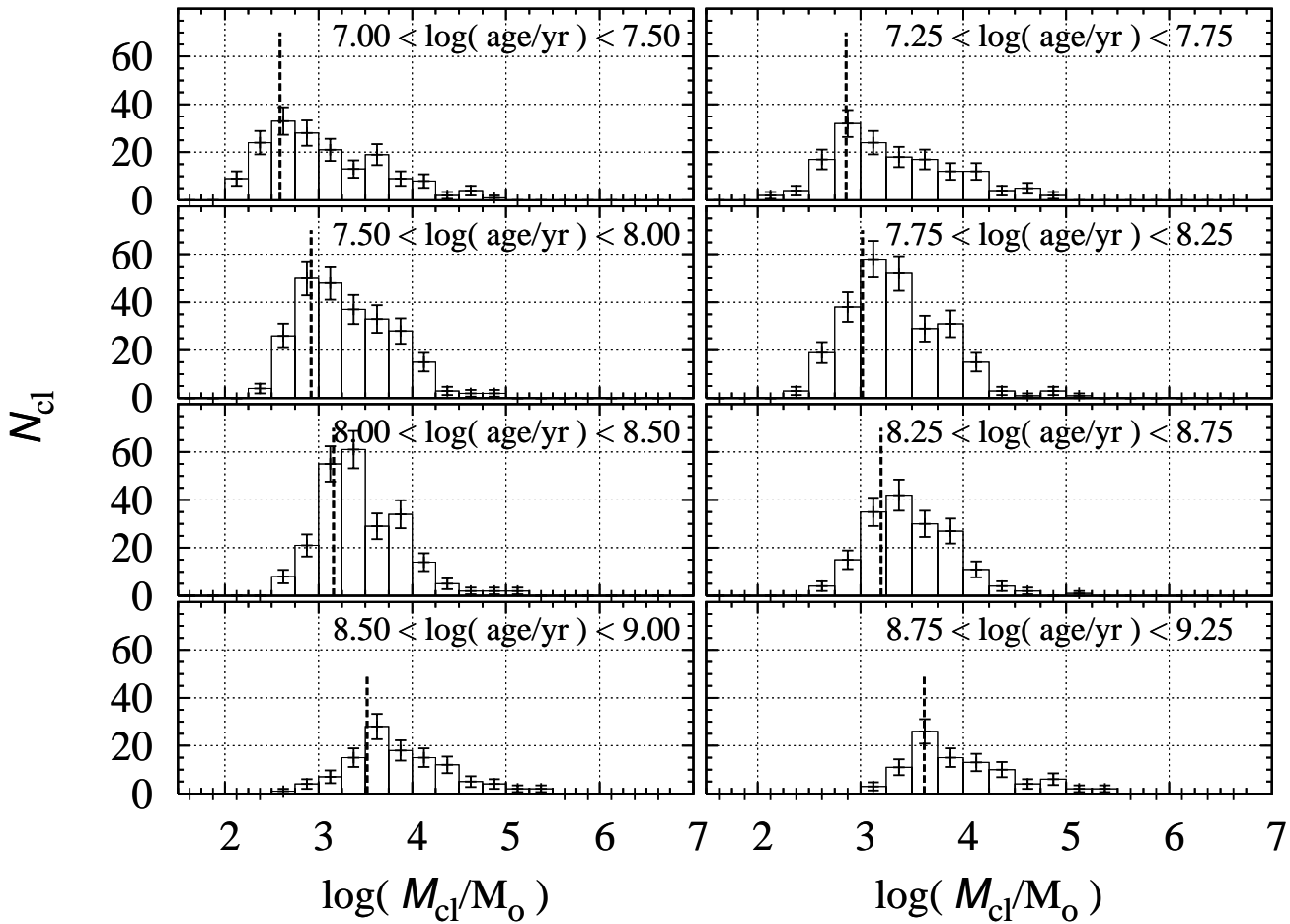, width=\linewidth}
\end{minipage}
\caption{Observed cluster mass functions for the age ranges included
at the top of each panel. In each panel, the vertical dashed line is
the mass limit bracketing 25 and 75 per cent of the cluster subsample
on either side. This is a good proxy to the cluster mass at the
turn-over of each CMF. For each age range, turn-over masses are
indicated in ($\log (\rm age),\log (M_{\rm cl})$) space as filled
triangles in Fig.~\ref{fig:obs_am}. The mass limits defined by the
dashed lines evolve with time following a line of constant luminosity,
at $M_V=-4.7$ mag. This implies that the decrease in cluster numbers
observed for each CMF below its turn-over mass (i.e. below the
vertical dashed line) is mostly driven by incompleteness effects}
\label{fig:compl} 
\end{figure*}

\section{Data}
\label{sec:data}

The basis for our detailed re-analysis of the Magellanic Clouds star
cluster systems is provided by the $UBVR$ broad-band spectral energy
distributions (SEDs) of H03, based on Massey's (2002) CCD survey of
the Magellanic Clouds.

In a series of recent papers, we developed a sophisticated tool for
star cluster analysis based on broad-band SEDs, {\sc AnalySED}, which
we tested extensively both internally (de Grijs et al. 2003b,c; Anders
et al. 2004) and externally (de Grijs et al. 2005), using both
theoretical and observed young to intermediate-age ($\lesssim 3 \times
10^9$ yr) star cluster SEDs, and the {\sc galev} ``simple'' stellar
population (SSP) models (Kurth et al. 1999; Schulz et al. 2002). The
accuracy for younger ages has since been increased via the inclusion
of an extensive set of nebular emission lines, as well as gaseous
continuum emission (Anders \& Fritze-v. Alvensleben 2003). We
concluded that the {\it relative} ages and masses within a given
cluster system can be determined to a very high accuracy depending on
the specific combination of passbands used (Anders et al. 2004). Even
when comparing the results of different groups using the same data
set, we can retrieve prominent features in the cluster age and mass
distributions to within $\Delta \langle \log( \mbox{age yr}^{-1} )
\rangle \le 0.35$ and $\Delta \langle \log( M_{\rm cl} / {\rm M}_\odot
) \rangle \le 0.14$, respectively (de Grijs et al. 2005), which
confirms that we understand the uncertainties associated with the use
of our {\sc AnalySED} tool to a very high degree.

In de Grijs \& Anders (2006) we presented newly and homogeneously
redetermined age and mass estimates for the entire Large Magellanic
Cloud (LMC) star cluster sample covered by the Massey (2002) data. Our
cluster age and mass determinations assume an average metallicity of
$Z = 0.008$ (where Z$_\odot = 0.020$); for the total extinction
towards the LMC clusters, we assumed $E(B-V) = 0.10$ mag, using the
Calzetti attenuation law (Calzetti 1997, 2001; Calzetti et al. 2000;
Leitherer et al. 2002) with $R_V = 4.05$. This corresponds to $E(B-V)
\simeq 0.13$ mag for both the Cardelli et al. (1989) and the Schlegel
et al. (1998) extinction laws. Based on the comparison of our results
in de Grijs \& Anders (2006) with those published previously in a
range of independent studies, and additionally on a detailed
assessment of the age-metallicity and age-extinction degeneracies, we
concluded that our broad-band SED fits yield reliable ages, with
statistical {\it absolute} uncertainties within $\Delta\log(
\mbox{age yr}^{-1}) \simeq 0.4$ overall.

The determination of the 50 per cent completeness limit of the LMC
cluster data was in essence based on a close inspection of H03's
fig. 4. These authors selected their sample from the existing
catalogues of Bica et al. (1999) and Pietrzy\'nski et al. (1999),
matched to the observational field of view of the Massey (2002)
data. Therefore, our completeness is that of these catalogues; Hunter
and her team did not quantify the completeness levels themselves
(D. Hunter, priv. comm.), although they discuss an observed fading
limit. However, for our analysis it is important to understand the
sample incompleteness affecting our observations (although this may
not be a strict ``limit'', but a ``range'' instead, given the nature
of the LMC cluster sample). As such, we adopted the conservative
approach that the present-day LMC cluster luminosity function (CLF;
see H03's fig. 4) is best represented by a power-law function in
luminosity. We obtained the best-fitting power law to the LMC's CLF
for $M_V \leq -4.5$ mag, where we assume the cluster sample to be
affected by negligible incompleteness. Based on this assumption, we
determined the 50 per cent completeness limit to occur at $M_V = -4.25
\pm 0.25$ mag, i.e., at level brigther by about 0.75 mag than H03's
observed fading limit. We note that if the underlying CLF is {\it not}
a power law (e.g., because of fading and secular evolution), the limit
we adopt following this approach is in fact a lower limit. In the
latter case the observations will likely be more complete than
estimated here.

In order to draw statistically robust conclusions from the
simulations performed in this paper, knowledge of the brightness limit
$M_V ^{\lim}$ above which the sample can be considered (fairly) complete
is required.  In section \ref{sec:synth}, we will derive an estimate
of $M_V ^{\lim}$ from a close inspection of the evolution with time of 
the observed CMF.  

\section{Building synthetic cluster populations above the fiducial detection limit}
\label{sec:synth}

In order to estimate the characteristic disruption time-scale of the
LMC clusters, we set up a grid of synthetic cluster populations
governed by various cluster formation histories, ICMFs and disruption
time-scales. The age and mass distributions of the surviving clusters
brighter than a given detection limit (see below) are then compared to
their observational counterparts in order to retrieve the best
possible constraints on the formation and evolution history of the LMC
star cluster system.

For the ICMF we assume a power-law mass spectrum ${\rm d}N/{\rm
d}M_{\rm cl} \propto M_{\rm cl}^{-2}$ over the mass range $10^2\,{\rm
M_{\odot}}-10^7\,{\rm M_{\odot}}$. Power-law cluster mass spectra with
slope $\simeq -2$ have been found for the young cluster systems in the
Antennae system (e.g., Zhang \& Fall 1999), the Whirlpool galaxy M51
(Bik et al.~2003) and for the Galactic open clusters (Battinelli et
al.~1994; Lada \& Lada 2003). 

Our knowledge of the cluster {\it formation} history of the LMC
remains rather poor at present. In order to explore the LMC's CFR,
star clusters are first distributed based on a constant cluster
formation rate. For each synthetic cluster population, the age
distribution of the surviving clusters that are more massive than the
detection limit is compared to that observed. From that comparison,
the CFR is corrected in order to match the modelled to the observed
age distribution. We will illustrate this step in Section
\ref{sec:disr}. Regarding the cluster age range, the cluster sample we
will first be focusing on spans the logarithmic age range $7.7 < \log
(\mbox{age yr}^{-1}) < 9.2$. Since our aim is to estimate the {\it
secular} cluster disruption time-scale, avoiding the first 50 Myr is
crucial, as discussed in Section \ref{sec:intro}. For the upper limit
to the age range we adopt $\log (\mbox{age yr}^{-1}) = 9.2$, since at
older ages our cluster sample contains very few objects. Once we have
a firm grasp on the cluster formation history for ages $7.7 <
\log(\mbox{age yr}^{-1}) < 9.2$, we will proceed one step further and
infer what the cluster formation history at older ages might have been
(see Section \ref{sec:agap}).

Now that we have defined the initial conditions of our fiducial
cluster populations in terms of their ages and masses, we have to
evolve each cluster mass over time. Lamers et al.~(2005b) showed that
the decreasing mass of a cluster can be described accurately as:
\begin{equation}
\frac{M_{\rm cl}(t)}{M_{\rm i}}=\left\{ \bigl[ \mu_{{\rm
ev}}(t)\bigr]^\gamma - \frac{\gamma t}{t^{\rm dis}}
\right\}^{1/\gamma}\,,
\label{cl_mass_evol_L.eq}
\end{equation}

where $M_{\rm cl}(t)$ is the mass of a cluster with initial mass
$M_{\rm i}$ that is still bound at an age $t$. In
Eq. \ref{cl_mass_evol_L.eq}, $\mu_{\rm ev}(t)$ is the fractional mass
decrease of the cluster because of stellar evolution only. The
temporal evolution of $\mu_{\rm ev}(t)$ is given by eqs. (2) and (3)
of Lamers et al. (2005b), which match the predictions of the {\sc
galev} ``simple'' stellar population models very accurately. 
As for the cluster disruption time-scale $t^{\rm dis}$, it scales with 
the disruption time-scale $t_4^{\rm dis}$ of a star cluster with an 
initial mass of $10^4 {\rm M}_\odot$ as
\begin{equation}
t^{\rm dis} =  t_4^{\rm dis} (M_{\rm i}/10^4 {\rm M}_\odot)^{\gamma}\,.
\label{eq:tdis}
\end{equation}
With $\gamma = 0.62$, these analytical descriptions of the evolution of the
mass of a given cluster matches the results of Baumgardt \& Makino
(2003), those being based on a large set of $N$-body simulations, where the
combined effects of stellar evolution, two-body relaxation and an
external tidal field were taken into account (see also Boutloukos \&
Lamers 2003 and Gieles et al.~2005 for empirical estimates of
$\gamma$; but see also Waters et al.~2006 and section \ref{sec:disr}
for another $\gamma$ value).  In what follows, the rate of cluster disruption 
is quantified by the estimate of $t_4^{\rm dis}$. 

Once we have evolved the cluster masses to the appropriate ages based
on the adopted CFR sampling, surviving clusters ending up below the
detection limit are excluded from the output sample. Because star
clusters fade with time as a result of stellar evolution, any
magnitude-limited cluster sample will be affected by an increasing
lower mass limit with increasing age, thus making it harder to detect
low-mass clusters at old age. Figure \ref{fig:obs_am} shows the cluster sample in
 log(mass) versus log(age) space, based on the analysis of de Grijs \&
Anders (2006). Although the lower cluster mass increases with
increasing age as expected, the distribution does not show a {\it
sharp} lower-mass cut-off. This is owing to the nature of the sample,
as it is in essence based on published cluster catalogues, each
affected by their own sampling characteristics. To establish whether
the cluster sample in Fig.~\ref{fig:obs_am} is in fact
magnitude-limited and, if so, to determine the magnitude limit above
which the sample is (fairly) complete, we must therefore resort to
empirical means.

Figure \ref{fig:compl} shows the cluster mass functions (i.e., the
number of clusters per logarithmic cluster mass interval) as a
function of age. Specifically, each panel of Fig.~\ref{fig:compl}
displays the mass function of a star cluster subsample covering
0.5\,dex in $\log({\rm age})$ in steps of 0.25\,dex, as indicated in
the panel legends. Each mass function exhibits a turn-over. If this
marks the age-dependent lower cluster mass limit below which
incompleteness affects the cluster sample significantly, then the
turn-over mass as a function of age must follow a trend parallel to
the detection limit in the ($\log({\rm age}),\log(M_{\rm cl})$) plane
(i.e. one of the dash-dotted lines in Fig.~\ref{fig:obs_am}). For
each of the cluster mass functions, the turn-over mass is
well-represented by the vertical thick dashed lines in the panels of
Fig.~\ref{fig:compl}. This is defined as the mass limit bracketing 25
and 75 per cent of the relevant cluster subsample. The evolution of
that mass limit with cluster age is shown as the triangles
(upright/upside-down triangles for the left/right-hand panels) in
Fig.~\ref{fig:obs_am}. We note that they follow closely a line of
constant magnitude, corresponding to $M_V = -4.7$ mag (thick
dash-dotted line with small squares). This result implies that our
star cluster sample is predominantly magnitude-limited. The decrease
in cluster numbers below the vertical line in each panel of
Fig.~\ref{fig:compl} (i.e., below the turn-over mass) is mostly driven
by incompleteness effects, affecting clusters fainter than $M_V =
-4.7$ mag. In the following, we adopt $M_V^{\rm lim} = -4.7$ mag as
the magnitude limit above which our sample is reasonably complete, 
referred to by the "fiducial detection limit".  Comparisons between
our model output and the photometric estimates of the cluster ages and
masses will be limited to star clusters of $M_V \le -4.7$ mag.  This 
represents a sample of 375 clusters, identified by crosses in Fig.~\ref{fig:obs_am}. 
In section \ref{sec:comp}, we will investigate how the estimate for the
{\dts} depends on this mass limit in ($\log({\rm age}),\log (M_{\rm
cl}$) space.

In summary, we have described how we evolve cluster populations drawn
from a given ICMF, and characterised by a given CFR and disruption
rate. We have also defined a criterion to separate the surviving
clusters of interest from those expected to be affected by significant
levels of incompleteness. We are now ready to compare these synthetic
populations to the observed cluster sample, represented by the crosses
in Fig.~\ref{fig:obs_am}, in terms of (i) the age distribution, (ii)
the mass distribution and (iii) the two-dimensional ($\log({\rm
age}),\log(M_{\rm cl})$) plane. To do so, both the observed and the
model distributions for the clusters in the age range $7.7 < \log
(\mbox{age yr}^{-1}) < 9.2$ and above the detection limit at $M_V^{\rm
lim} = -4.7$ mag are binned identically. We adopt bins of 0.3\,dex in
both $\log({\rm age})$ and $\log (M_{\rm cl})$. The corresponding
distribution of cells in ($\log({\rm age}),\log(M_{\rm cl})$) space is
shown as the grid in Fig.~\ref{fig:obs_am}. In the next section, we
will make sure that the choice of the bin size does not affect our
results. Given the presence of Poissonian statistics in both the data
and the modelling, the goodness-of-fit is quantified using the Poisson
Probability Law ({\it PPL}) introduced by Dolphin (2002) (see also
Dolphin \& Kennicutt 2002). This follows
\begin{equation}
{\it PPL} = 2 \sum_{i=1}^{N} \left( m_i - n_i + n_i {\rm ln}
\frac{n_i}{m_i} \right)\,,
\end{equation}
where $N$ is the number of cells, $m_i$ the predicted number of
clusters in cell $i$ and $n_i$ its observed counterpart. The {\it PPL}
estimator is the Poissonian equivalent of the $\chi^2$ estimator,
i.e., lower values correspond to a better description of the data.
Once divided by the number of bins minus the number of degrees of
freedom, we obtain the reduced {\it PPL}, which we refer to as
$\chi_{\nu}^2$.

\section{Cluster disruption time-scale}
\label{sec:disr}

As mentioned in Section \ref{sec:intro}, to infer an estimate (or a
range of estimates, as we will see below) for the {\dts} is equivalent
to deducing the temporal evolution of the LMC's CFR over the past
1.5\,Gyr (i.e., up to the upper age limit we consider at the present
stage of our analysis).

\begin{figure}
\begin{minipage}[t]{\linewidth}
\centering\epsfig{figure=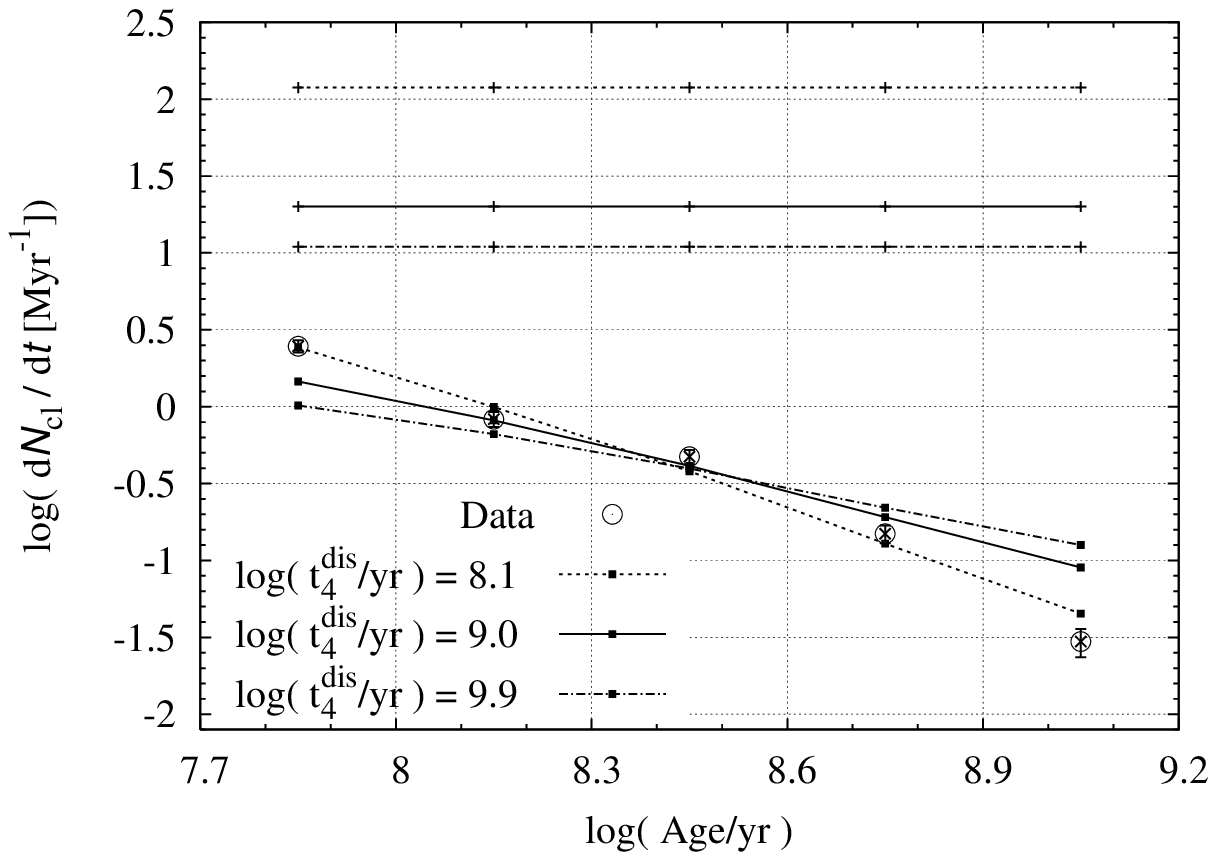, width=\linewidth} 
\end{minipage}
\vfill
\begin{minipage}[t]{\linewidth}
\centering\epsfig{figure=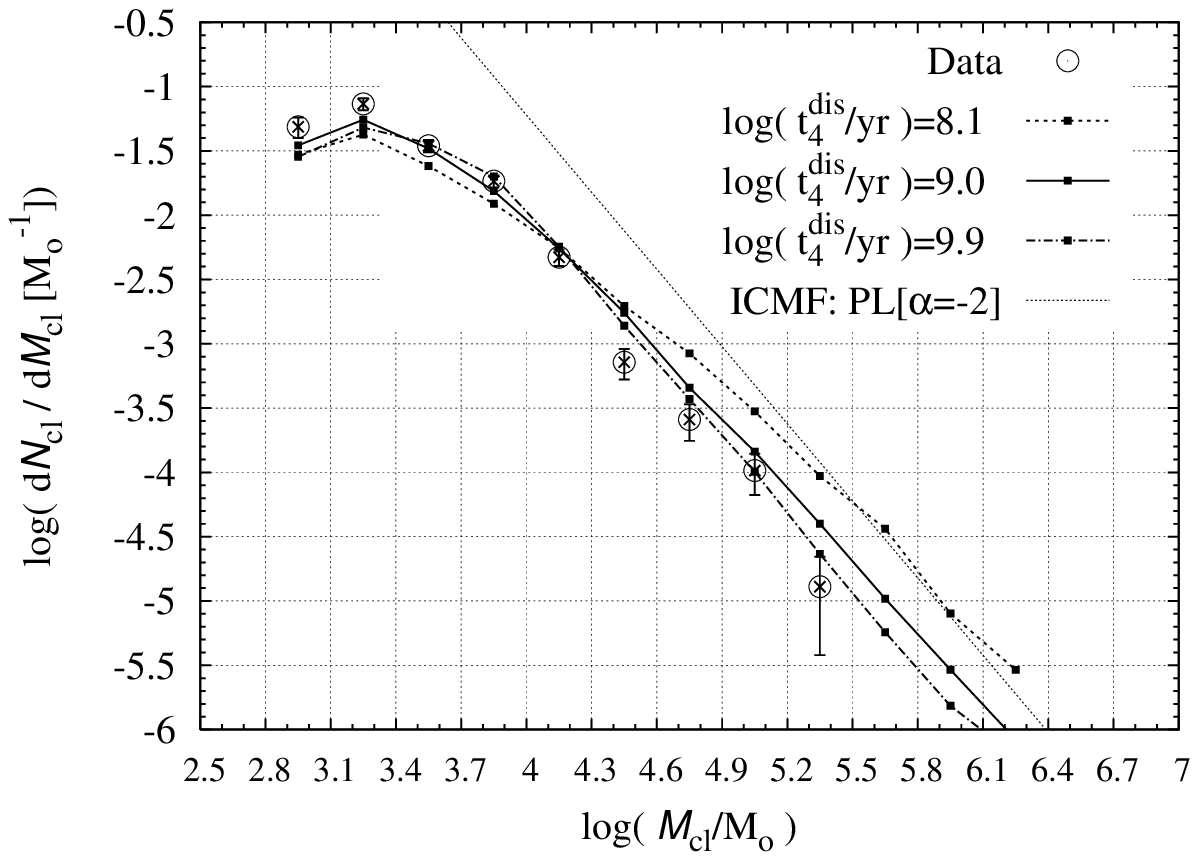, width=\linewidth}
\end{minipage}
\caption{{\it Top:} Age distribution of the LMC star clusters brighter
than the fiducial detection limit at $M_V^{\rm lim}=-4.7$ mag
(i.e. more massive than the thick dash-dotted line in
Fig.~\ref{fig:obs_am}) integrated over mass (large circles with error
bars). The lower curves with small squares are their modelled
counterparts for three cluster disruption time-scales, $\log (t_4^{\rm
dis} {\rm yr}^{-1})=8.1, 9.0$ and $9.9$. They have been shifted
vertically in order to contain 375 clusters as observed. The upper horizontal straight
lines are the corresponding CFRs (assumed constant here). {\it
Bottom:} Same for the evolved cluster mass spectra (i.e., the number
of clusters per constant linear mass interval) integrated over
age. The dotted straight line shows the slope $\alpha = -2$ of 
the power-law initial cluster mass spectrum.}
\label{fig:cstcfr} 
\end{figure}

\begin{figure}
\begin{minipage}[t]{\linewidth}
\centering\epsfig{figure=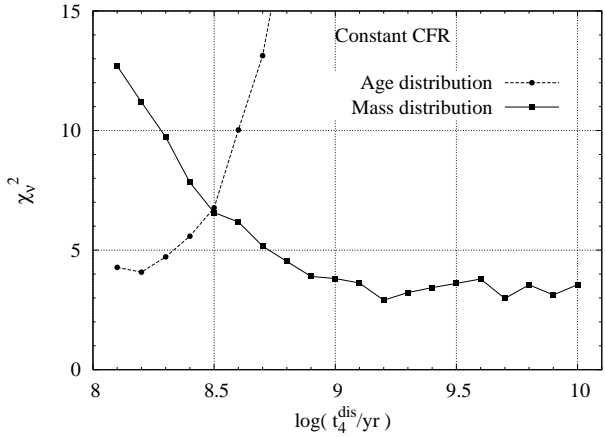, width=\linewidth}
\end{minipage}
\caption{Reduced $\chi^2_{\nu}$ as a function of the characteristic
disruption time-scale, $t_4^{\rm dis}$, of a $10^4\,{\rm M}_{\odot}$
cluster, as inferred from the age distribution integrated over mass
and the mass distribution integrated over age of the detected
surviving clusters. Clusters are assumed to have formed with a
constant CFR between $\log (\mbox{age yr}^{-1})=7.7$ and
$\log(\mbox{age yr}^{-1})=9.2$.}
\label{fig:xhi2_cst} 
\end{figure}

First, we simulated the evolution of 20 putative star cluster systems.
They are all characterized by a constant CFR over the age range $7.7 <
\log(\mbox{age yr}^{-1}) < 9.2$ and have the same power-law ICMF of 
spectral index $\alpha =-2$. 
The cluster systems differ with respect to their {\dts}s, for
which twenty values were tested, i.e., $\log (t_4^{\rm dis} {\rm
yr}^{-1})=8.1$ to 10.0 in steps of 0.1 dex.  
We then normalise the evolved cluster age/mass distributions so
that they contain a constant number of 375 clusters, to match the
observations. The initial distributions (e.g., the CFR) follow from
this. As a result, a shorter {\dts} implies a higher CFR, in order to
maintain the present number of observed clusters despite more
vigourous disruption. In practice, the CFR for $\log (t_4^{\rm dis}
{\rm yr}^{-1})=8.1$ is ten times higher than for $\log (t_4^{\rm dis}
{\rm yr}^{-1})=9.9$ (see top panel of Fig.~\ref{fig:cstcfr}).

When the {\dts} is long (e.g., $\log (t_4^{\rm dis} {\rm yr}^{-1}) =
9.9$), the decrease in the number of evolved clusters per time unit as
a function of increasing age is dominated by fading effects. The slope
of the age distribution therefore mirrors the slope $\zeta \simeq 0.7$
of the lower mass limit over time in ($\log({\rm age}), \log(M_{\rm
  cl})$) space and is $\zeta (1 + \alpha) \simeq -0.7$ ($\alpha \simeq
-2$). In the opposite case, when the {\dts} is short (e.g., $\log
(t_4^{\rm dis} {\rm yr}^{-1})=8.1$), the slope of the age distribution
is mostly determined by disruption and follows $(1 - \alpha)/\gamma
\simeq -1.6$ (see Boutloukos \& Lamers 2003 for an in-depth
discussion).

From the observed age distribution alone, one may deduce that the
characteristic disruption time-scale of a $10^4 {\rm M}_\odot$ cluster
in the LMC is close to 100 Myr. Shown graphically, in the top panel of
Fig.~\ref{fig:cstcfr}, the lower dotted curve (corresponding to the
age distribution for $\log (t_4^{\rm dis} {\rm yr}^{-1})=8.1$) follows
the data points better than the dash-dotted curve (i.e., the age
distribution for $\log (t_4^{\rm dis} {\rm yr}^{-1})=9.9$). In the
latter case, the predicted age distribution is shallower (with a slope
$\simeq -0.7$) than the observed cluster age distribution (exhibiting
a slope $\lesssim -1.6$). This conclusion is also supported by the
relation between $t_4^{\rm dis}$ and $\chi_{\nu}^2$ for the age
distribution (i.e., the long-dashed curve with the filled circles in
Fig.~\ref{fig:xhi2_cst}), which shows a steeply rising $\chi_{\nu}^2$
for $t_4^{\rm dis} > 200$ Myr.

\begin{figure}
\begin{minipage}[t]{\linewidth}
\centering\epsfig{figure=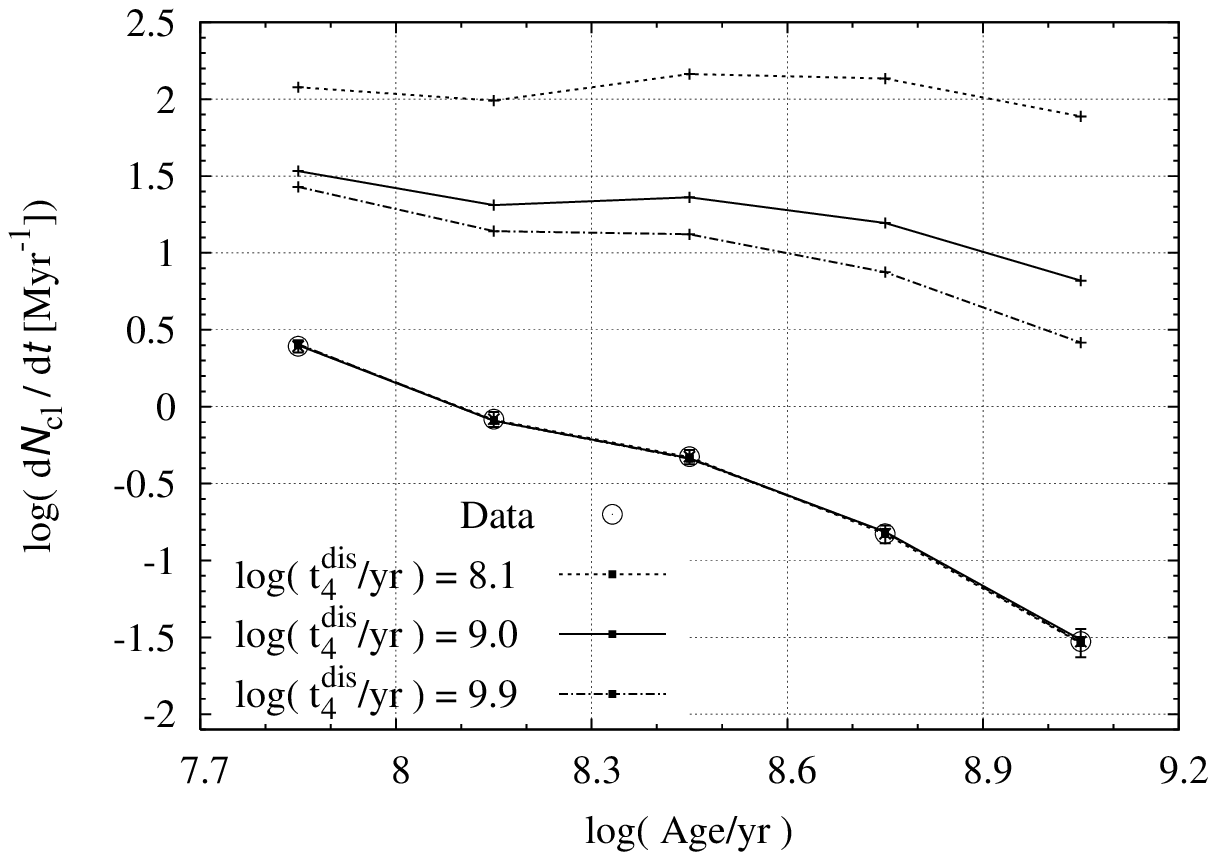, width=\linewidth} 
\end{minipage}
\vfill
\begin{minipage}[t]{\linewidth}
\centering\epsfig{figure=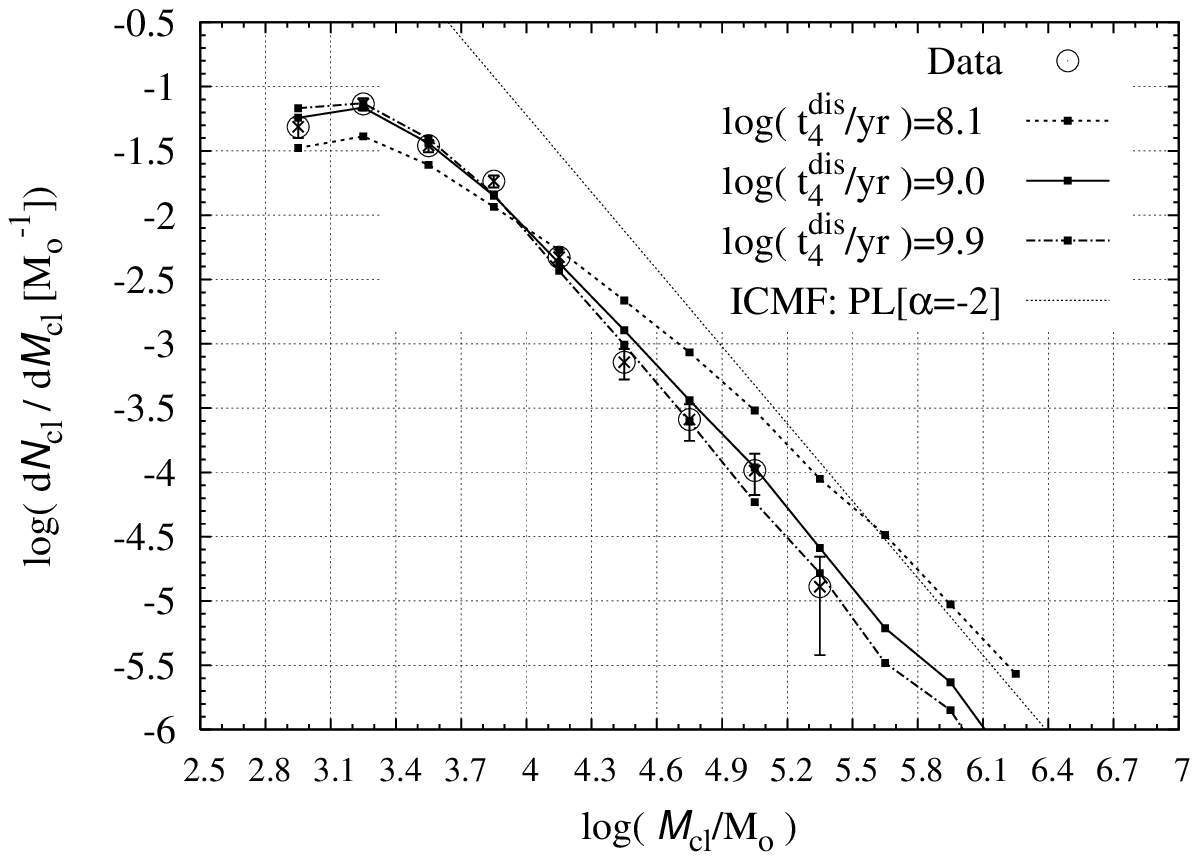, width=\linewidth}
\end{minipage}
\caption{Same as Fig.~\ref{fig:cstcfr} except for the CFR, which in
this case is adjusted in order to match the model age distribution to
the observed distribution for each cluster disruption time-scale. {\it
Top:} Cluster age distributions {\it Bottom:} Cluster mass spectra}
\label{fig:varcfr} 
\end{figure}

\begin{figure}
\begin{minipage}[t]{\linewidth}
\centering\epsfig{figure=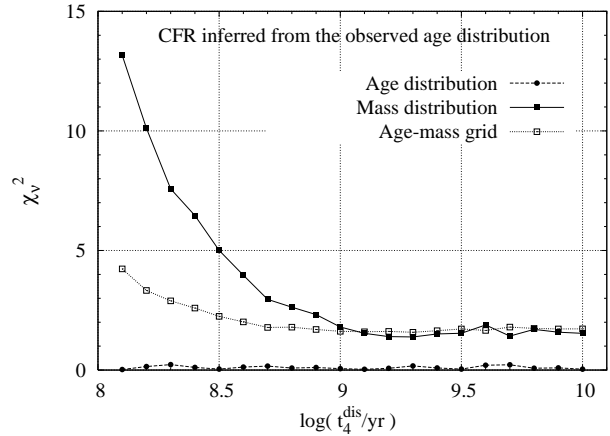, width=\linewidth}
\end{minipage}
\caption{Reduced $\chi^2_{\nu}$ as a function of $t_4^{\rm dis}$ for
model cluster age distributions reproducing the observed distribution
via CFR adjustements (top panel of
Fig.~\ref{fig:varcfr}). $\chi^2_{\nu}$ functions are shown for the
cluster age distribution (equal to zero but for size-of-sample related
Poissonian noise), for the cluster mass spectrum and for the
two-dimensional distribution of clusters in ($\log({\rm age}),\log(M_{\rm
cl})$) space. }
\label{fig:xhi2_var} 
\end{figure}

The ($t_4^{\rm dis}$, $\chi_{\nu}^2$) relation for the cluster mass
spectrum integrated over age paints a completely different picture,
however. The short {\dts} derived from the age distribution alone now
results in a four times larger $\chi_{\nu}^2$ than for $t_4^{\rm dis}
\gtrsim 1$\,Gyr (the solid line with filled squares in
Fig.~\ref{fig:xhi2_cst}). The bottom panel of Fig.~\ref{fig:cstcfr}
illustrates that if $\log (t_4^{\rm dis} {\rm yr}^{-1}) = 8.1$, the
evolved cluster mass spectrum above $10^4\,{{\rm M}_{\odot}}$ is
significantly shallower than the observed CMF, of which the slope
remains close to that of the ICMF in that mass range. Better fits to
the evolved CMF are obtained for $\log (t_4^{\rm dis} {\rm
yr}^{-1})=9.0$ and 9.9. This discrepancy between two different
estimators for the model goodness-of-fit, $\chi_{\nu}^2$ from the age
distribution and $\chi_{\nu}^2$ from the mass distribution, suggests
an inconsistent hypothesis, presumably the assumption of a constant CFR.

It is worth emphasising that a possibly variable CFR in the LMC should
not be considered as an obstacle to the derivation of the cluster
disruption time-scale. Any sufficiently old, coeval subsample of
clusters is suitable to infer it, as long as the imprint of the
secular dynamical evolution can be detected. In fact, the limited age
range of a coeval population ensures that temporal CFR variations can
safely be ignored. The imprint of the secular dynamical evolution
leads to a turn-over in the CMF. Provided that the shape of the ICMF
is known, the disruption time-scale for the relevant environmental
conditions can be derived from the combination of the mean cluster age
and of the cluster mass at the CMF turn-over (see Section
\ref{sec:1vs10} and Fig.~\ref{fig:tevol_MSp}). This argument is
supported by our recent study of the roughly coeval intermediate-age
cluster population in M82 B (de Grijs, Parmentier \& Lamers 2005;
but see also Smith et al.~2007 for updated M82 B data).

In the following, we build on this principle. Specifically, we split
the cluster age range under consideration, $7.7 < \log(\mbox{age
yr}^{-1}) < 9.2$, into five subranges, each with a width of $\Delta
\log(\mbox{age yr}^{-1}) = 0.3$ (i.e., the size of a cell in
Fig.~\ref{fig:obs_am}). We assume that these age bins are narrow
enough so that each can be characterised by a single CFR. For each of
the twenty disruption time-scales explored, we first assume a constant
CFR and we derive the corresponding age distribution of the observed
survivors. From the comparison between the modelled and the observed
age distributions the CFR is adjusted so that the distributions match
each other. This is illustrated in the top panel of
Fig.~\ref{fig:varcfr}. Varying the {\dts} now leads to distinct
cluster formation histories, not only with respect to the mean CFR,
but also with respect to its temporal variations. If the {\dts} is
$\log (t_4^{\rm dis}{\rm yr}^{-1})=8.1$, the CFR is almost constant
and of order 100 clusters per Myr. This is consistent with what we
concluded above (see the top panel of Fig.~\ref{fig:cstcfr}). From
that same figure, we also found that if the CFR is constant and if
$\log (t_4^{\rm dis}{\rm yr}^{-1}) = 9.0$ or 9.9, the derived age
distribution is shallower than observed. As a result, for such a long
{\dts}, the LMC cluster age distribution can be matched only by
adopting a CFR that decreases with increasing age (compare the top
panels of Figs.~\ref{fig:cstcfr} and \ref{fig:varcfr}). Once an
appropriate CFR has been adopted for each age bin, the construction of
a consistent integrated cluster mass distribution follows naturally
(see the bottom panel of Fig.~\ref{fig:varcfr}).

\begin{table}
\caption[]{Summary of the various Poissonian $\chi^2$ tests. The first
column describes the diagnostic used to compare the model to the
data, either the mass function or the age-mass grid.  For the fiducial case, we also include the results
for each of the five mass functions per age bin, with each age range
quoted as [min(log(age yr$^{-1}$)),Max(log(age yr$^{-1}$))]. The
second column tabulates the minimum $\chi^2_{\nu}$.  The third column
contains, first, $\log (t_4^{\rm dis} {\rm yr}^{-1})$ at which
$\chi^2_{\nu}$ is minimum, and secondly the range of $\log (t_4^{\rm
dis} {\rm yr}^{-1})$ for which $\chi^2_{\nu} \le \chi^2_{\nu ,min}+1 $. 
The fourth column refers to the relevant figure.
Results are provided for the fiducial case, for other detection limits
corresponding to the three thin dash-dotted lines in
Fig.~\ref{fig:obs_am} (referred to by the amplitude of the vertical
shift in $\log(M_{\rm cl})$ with respect to the fiducial detection limit; 
see section \ref{sec:comp}), for ICMFs with different slopes ($\alpha
=-1.9$ and $-2.1$) and for a different model of cluster mass-loss 
($\gamma =1$ in equations \ref{cl_mass_evol_L.eq} and~\ref{eq:tdis}).}
\label{tab:xhi2}
\begin{tabular}{  l l c l  } \hline 
   & $\chi^2_{\nu ,min}$ & $\log (t_4^{\rm dis} {\rm yr}^{-1})$ &  \\ \hline

\multicolumn{4}{c}{Fiducial Case} \\ \hline
Mass function & $\simeq 1.5$ & $(\geq 9.10, \geq 8.85)$    & Fig.~\ref{fig:xhi2_var} \\
Age-mass grid & $\simeq 1.7$ & $(\geq 8.70, \geq 8.35)$    & Fig.~\ref{fig:xhi2_var} \\
CMF [7.7-8.0] & $\simeq 2.5$ & $(8.70, {\rm any})$   & Fig.~\ref{fig:xhi2_MFagebin} \\
CMF [8.0-8.3] & $\simeq 1.3$ & $(\geq 9.00, \geq 8.55)$      & Fig.~\ref{fig:xhi2_MFagebin} \\
CMF [8.3-8.6] & $\simeq 2.0$ & $(\geq 9.40, \geq 9.0)$       & Fig.~\ref{fig:xhi2_MFagebin} \\
CMF [8.6-8.9] & $\simeq 0.9$ & $(9.00, [8.5,9.7] )$ & Fig.~\ref{fig:xhi2_MFagebin} \\
CMF [8.9-9.2] & $\simeq 0.8$ & $(\geq 9.2, \geq 8.7 )$      & Fig.~\ref{fig:xhi2_MFagebin} \\ \hline

\multicolumn{4}{c}{$\alpha = -1.9$}   \\ \hline 
Mass function      & $\simeq 1.7$ & $(\geq 9.40, \geq 9.00)$    & Fig.~\ref{fig:rob_xhi2} \\
Age-mass grid      & $\simeq 1.6$ & $(\geq 8.90, \geq 8.45)$    & Fig.~\ref{fig:rob_xhi2} \\
\hline

\multicolumn{4}{c}{$\alpha = -2.1$ } \\ \hline
Mass function     & $\simeq 1.4$ & $(\simeq 9.30, \geq 8.65)$    & Fig.~\ref{fig:rob_xhi2}  \\
Age-mass grid     & $\simeq 1.6$ & $(\simeq 8.80, \geq 8.25)$    & Fig.~\ref{fig:rob_xhi2}   \\
\hline

\multicolumn{4}{c}{$\gamma = 1.00$ } \\ \hline
Mass function     & $\simeq 1.4$ & $(\geq 9.20, \geq 9.05)$    & Fig.~\ref{fig:rob_xhi2}  \\
Age-mass grid     & $\simeq 1.7$ & $(\geq 8.90, \geq 8.60)$    & Fig.~\ref{fig:rob_xhi2}   \\
\hline

\multicolumn{4}{c}{$\Delta \log(M_{\rm cl}) = +0.20$ (line `[1]' in Fig.~\ref{fig:obs_am}) } \\ \hline
Mass function      & $\simeq 1.6$ & $(\geq 9.10, \geq 8.70)$    & Fig.~\ref{fig:rob_xhi2_dct}  \\
Age-mass grid      & $\simeq 2.0$ & $(\geq 8.70, \geq 8.45)$    & Fig.~\ref{fig:rob_xhi2_dct} \\ \hline

\multicolumn{4}{c}{$\Delta \log(M_{\rm cl}) = -0.20$ (line `[2]' in Fig.~\ref{fig:obs_am})} \\ \hline
Mass function      & $\simeq 3.0$ & $(\simeq 9.00, [8.70,9.50])$    & Fig.~\ref{fig:rob_xhi2_dct} \\
Age-mass grid      & $\simeq 2.0$ & $(\geq 9.00, \geq 8.45)$    & Fig.~\ref{fig:rob_xhi2_dct} \\ \hline

\multicolumn{4}{c}{$\Delta \log(M_{\rm cl}) = -0.48$ (line `[3]' in Fig.~\ref{fig:obs_am})} \\ \hline
Mass function      & $\simeq 8.4$ & $(\geq 8.65, [8.45,8.90])$    & Fig.~\ref{fig:rob_xhi2_dct} \\
Age-mass grid      & $\simeq 3.1$ & $(\geq 8.60, [8.35,9.30])$    & Fig.~\ref{fig:rob_xhi2_dct} \\ \hline

\end{tabular}
\end{table}  

\begin{figure*}
\begin{minipage}[t]{\linewidth}
\centering\epsfig{figure=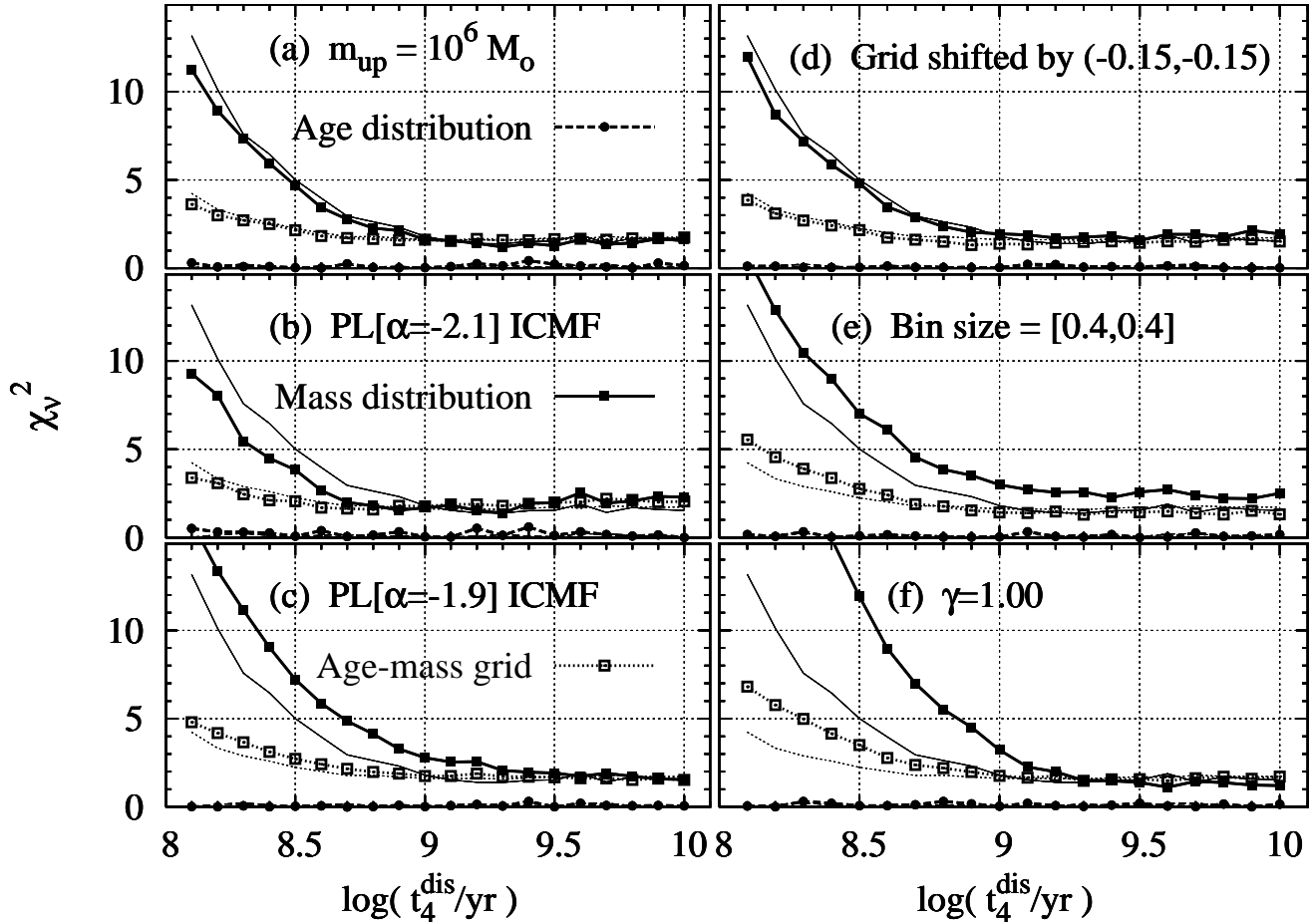, width=\linewidth}
\end{minipage}
\caption{Evolution of the reduced $\chi^2_{\nu}$ vs. {\dts} $t_4^{\rm
dis}$ for seven different models applied to the data. The fiducial case
(upper limit of the cluster mass distribution $M_{\rm cl,up} =
10^7\,{\rm M}_{\odot}$, ICMF spectral index $\alpha =-2$, grid location 
and cell size as shown in Fig.~\ref{fig:obs_am}, $\gamma =0.62$) is shown as the thin 
symbol-free lines in each panel for the sake of comparison.  
The subsequent panels consider: 
{\it (a)} a different upper mass limit of $M_{\rm cl,up} = 10^6\,{\rm M}_{\odot}$,
{\it (b)} a steeper power-law ICMF with $\alpha=-2.1$, 
{\it (c)} a shallower power-law ICMF with $\alpha=-1.9$, 
{\it (d)} a grid location shifted by $-0.15$\,dex in both $\log({\rm age})$ and
$\log(M_{\rm cl})$, 
{\it (e)} a grid cell size of $\Delta \log(M_{\rm cl}) = 0.4$ and 
$\Delta \log({\rm age}) = 0.4$ (instead of 0.3 in the fiducial case), and
{\it (f)} a larger value of the exponent $\gamma$ describing the dependence of the 
disruption timescale on the initial cluster mass 
(see equations \ref{cl_mass_evol_L.eq} and \ref{eq:tdis}), respectively.  
The corresponding $\chi^2_{\nu}$ functions are
shown as the thick lines with symbols}
\label{fig:rob_xhi2} 
\end{figure*}

The relation between the reduced $\chi_{\nu}^2$ and the {\dts}
$t_4^{\rm dis}$ is presented in Fig.~\ref{fig:xhi2_var}. Since the
temporal variations in the CFR are adjusted to reproduce the LMC
cluster age distribution, the $\chi_{\nu}^2$ for the cluster age
distribution is close to zero, regardless of the adopted disruption
time-scale. Its best estimate must now be inferred from the
$\chi_{\nu}^2$ for the cluster mass spectrum (the solid curve with the
filled squares). The best fits are obtained for 
$\log(t_4^{\rm dis} {\rm yr}^{-1}) \ge 9.0$, for which $\chi_{\nu}^2$ is minimum at $\simeq
1.5$. As for a constant CFR, the shortest tested {\dts} ($\log
(t_4^{\rm dis} {\rm yr}^{-1})=8.1$) leads to an evolved mass spectrum
that is significantly shallower than observed (see the short-dashed
curve in the bottom panel of Fig.~\ref{fig:varcfr}).

In Fig.~\ref{fig:xhi2_var}, we also show the $\chi_{\nu}^2$ for the
distribution of detected clusters in ($\log({\rm age}),\log(M_{\rm
cl})$) space.  The constraints on $t_4^{\rm dis}$
derived from the two-dimensional distribution of data points are
significantly looser than those obtained from the mass spectrum (see
Table \ref{tab:xhi2}). This likely results from the smaller number of
clusters in each grid cell than in the mass distribution bins, leading
to larger error bars and hence more poorly determined constraints (see
also Section \ref{sec:1vs10}).  From the results shown in
Fig.~\ref{fig:xhi2_var}, we conclude that the best constraint we can
set on the characteristic {\dts} in the LMC is $t_4^{\rm dis} \geq
1$\,Gyr. As we will discuss in Section \ref{sec:1vs10}, the presently
available data do not allow us to distinguish between $t_4^{\rm dis} =
1$ and 10 Gyr.

Table \ref{tab:xhi2} lists the cluster 
disruption time-scales (or a range of cluster
disruption time-scales where relevant) corresponding to the minimum
$\chi _{\nu}^2$ for the cases discussed in this section, as
well as those presented in Section \ref{sec:comp}.
Also given is the range of accepted cluster disruption time-scales
defined as the most extreme models which satisfy
$\chi^2_{\nu} \le \chi^2_{\nu ,min}+1$.

We now test how robust the $\chi^2_{\nu}$ functions are with respect
to our input parameters. Fig.~\ref{fig:rob_xhi2} shows that they
remain practically unaffected when the adopted ICMF upper limit is
lowered from $M_{\rm cl,up} = 10^7\,{\rm M}_{\odot}$ down to $M_{\rm
cl,up} = 10^6\,{\rm M}_{\odot}$ (panel {\it (a)}), when the grid in
($\log({\rm age}),\log(M_{\rm cl})$) space is shifted by $-0.15$\,dex
in both $\log({\rm age})$ and $\log(M_{\rm cl})$ (panel {\it (d)}) and
when the size of the grid cells is $\Delta \log(M_{\rm cl}) = 0.4$ and
$\Delta \log({\rm age}) = 0.4$ dex (instead of 0.3 dex,panel {\it (e)}). 
In all panels, these three cases are illustrated by the thick curves, 
along with thin curves representing the fiducial case (from
Fig.~\ref{fig:xhi2_var}).  The $t_4^{\rm dis}$ ranges over which $\chi^2_{\nu}$ is
minimum is practically the same as for the fiducial case. Our
conclusions thus prove robust with respect to these variations.  

Thus far, we have assumed a power-law initial cluster mass spectrum
with a slope of $\alpha =-2$ for our
synthetic star cluster systems.  Slightly shallower or steeper power 
laws have been reported in the literature for observed CMFs, however, 
with $-2.1 \lesssim \alpha \lesssim -1.8$ (see de Grijs et al.~2003c for a review). 
In panels (b) and (c) of Figure \ref{fig:rob_xhi2}, we substituted the 
canonical ICMF with $\alpha =-2$ by a steeper power law of slope 
$\alpha =-2.1$ and a shallower power law of slope $\alpha =-1.9$, respectively. 
A steeper/shallower power law leads to larger/smaller ranges of
acceptable $t_4^{\rm dis}$, although changes are not significant (see also Table
\ref{tab:xhi2}). Therefore, we conclude that our estimate of the
characteristic cluster disruption time-scale in the LMC is also robust
with respect to changes in the ICMF, provided that this remains
consistent with the observed mass function of bound clusters at young age.

While Boutloukos \& Lamers (2003) and Gieles et al.(2005) advocate
that the cluster disruption timescale goes as the cluster mass to the power 
$\gamma =0.62$, Fall \& Zhang (2001) come up with $\gamma =1$ 
(see equations \ref{cl_mass_evol_L.eq} and~\ref{eq:tdis}).  Moreover,
in a recent study of the M87 globular cluster system, Waters 
et al.~(2006) find that a cluster evaporation model with $\gamma =1$ 
provides a better match to the observed cluster mass function than if $\gamma =0.62$
(see also the discussion in Gieles 2007).  We thus now assess
how much the uncertainties affecting $\gamma$ propagate as uncertainties
on $t_4^{\rm dis}$.  The bottom-right panel of Fig.~\ref{fig:rob_xhi2}
(panel (f)) compares between $\gamma =1$ (thick curves) and $\gamma =0.62$
(fiducial case; thin curves).  A greater value of $\gamma$ sets tighter constraints
on the cluster disruption timescale in the sense that the solution 
$t_4^{\rm dis} \le 1$\,Gyr is significantly more rejected.
A larger $\gamma$ actually leads to shallower cluster
disruption limits in log(mass) versus log(age) space 
(thick dashed lines in Fig.~\ref{fig:obs_am}).  That is,
for a fixed value of $t_4^{\rm dis}$, clusters initially more/less massive 
than $10^4\,{\rm M}_{\odot}$ dissolve more slowly/quickly if $\gamma =1$ than
if $\gamma =0.62$.  Owing to the combination of the adopted age range with the
fiducial detection limit, clusters less massive than $10^4\,{\rm M}_{\odot}$ dominate
the cluster sample in terms of number.  As a result, $\gamma =1$ mimics a quicker 
evolutionary rate for the overall cluster population
(even though $t_4^{\rm dis}$ is kept unchanged).
Therefore, to account either for the {\it given} observed cluster mass distribution 
or for the {\it given} cluster sample in ($\log({\rm age}),\log(M_{\rm cl})$) space 
requires a slower cluster disruption timescale when $\gamma =1$, as illustrated by 
panel (f) in Fig.~\ref{fig:rob_xhi2}.  The overall conclusions drawn on the basis
of the fiducial case, i.e. $t_4^{\rm dis} \gtrsim 1$\,Gyr and temporal variations 
of the CFR as shown on top panel of Fig.~\ref{fig:varcfr}, are not affected, however
(see also Table \ref{tab:xhi2}).            

How the $\chi^2_{\nu}$ functions depend on variations in the location of
the detection limit in ($\log({\rm age}),\log(M_{\rm cl})$) space will 
be discussed in Section \ref{sec:comp}. \\

\begin{figure}
\begin{minipage}[t]{\linewidth}
\centering\epsfig{figure=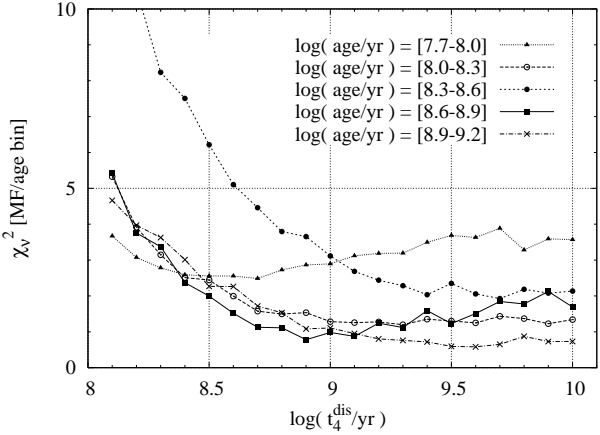, width=\linewidth}
\end{minipage}
\caption{Evolution of $\chi^2_{\nu}$ with {\dts} $t_4^{\rm dis}$ for
the CMFs for each age bin of the fiducial case. The age ranges are
listed in the figure legend; they correspond to the age binning used
in Fig.~\ref{fig:obs_am}}
\label{fig:xhi2_MFagebin} 
\end{figure}

As a way of inspecting more closely the goodness-of-fit of the
modelled grid to the data grid, we consider (in
Fig.~\ref{fig:xhi2_MFagebin}) the $\chi^2_{\nu}$ functions for each
CMF {\it for each age bin}. These are listed in the figure legend.
All cluster age ranges, except for the first one, support the same
conclusion, i.e., $t_4^{\rm dis} \gtrsim 1$\,Gyr (details are provided
in Table \ref{tab:xhi2}). The tightest constraint is provided by the
CMF in the age range $8.3 < \log(\mbox{age yr}^{-1}) < 8.6$
($\chi^2_{\nu}$ is minimum over the more limited age range $\log
(t_4^{\rm dis} {\rm yr}^{-1}) \ge 9.4$), which is also one of the two
most populated age bins.

The $\chi^2_{\nu}$ function for the age range $7.7 < \log(\mbox{age
yr}^{-1}) < 8.0$ (i.e., the youngest age bin) appears different.
At such a young age, the CMF bears the imprint of dynamical evolution
only if this proceeds on a short time-scale (e.g., $\log (t_4^{\rm
dis} {\rm yr}^{-1}) \simeq 8.1$), which is not the case here. 
We therefore expect a $\chi^2_{\nu}$ function that is flat over most of
the disruption time-scale range, except for a possible rise for the very
shortest time-scales.  Actually, the $\chi^2_{\nu}$ function is roughly 
constant over the two-decade $t_4^{\rm dis}$ variations. 

\begin{figure}
\begin{minipage}[t]{\linewidth}
\centering\epsfig{figure=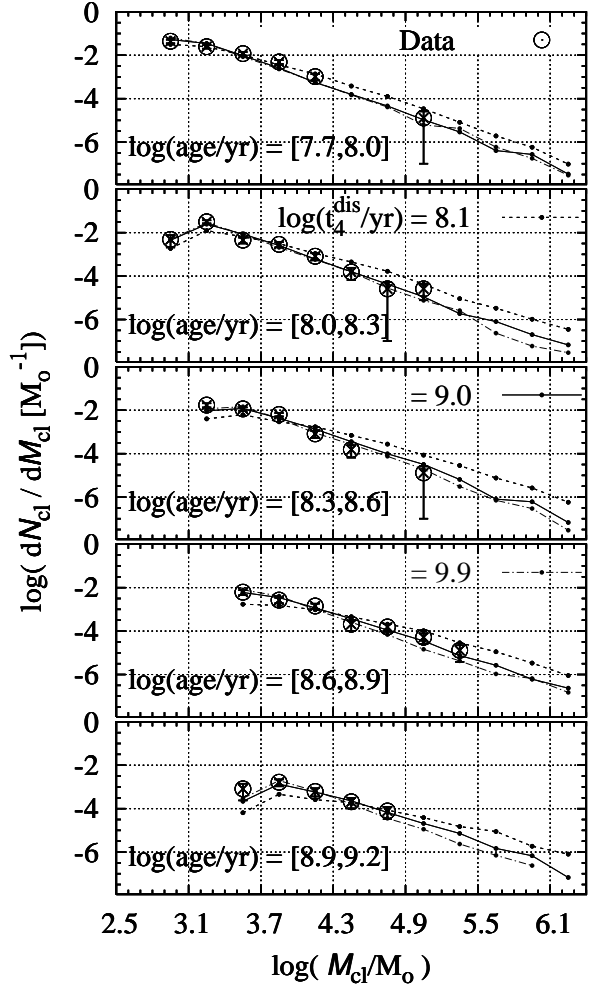, width=\linewidth}
\end{minipage}
\caption{Mass spectra of the detected surviving clusters for the data
(large open circles with error bars) and the models (short-dashed
lines: $\log (t_4^{\rm dis} {\rm yr}^{-1})=8.1$; solid lines: $\log
(t_4^{\rm dis} {\rm yr}^{-1})=9.0$; dash-dotted lines:$\log (t_4^{\rm
dis} {\rm yr}^{-1})=9.9$) for the age ranges considered in
Fig.~\ref{fig:xhi2_MFagebin}. They correspond to the age binning used
in Fig.~\ref{fig:obs_am}}
\label{fig:MSp_ab} 
\end{figure}

\begin{figure*}
\begin{minipage}[t]{\linewidth}
\centering\epsfig{figure=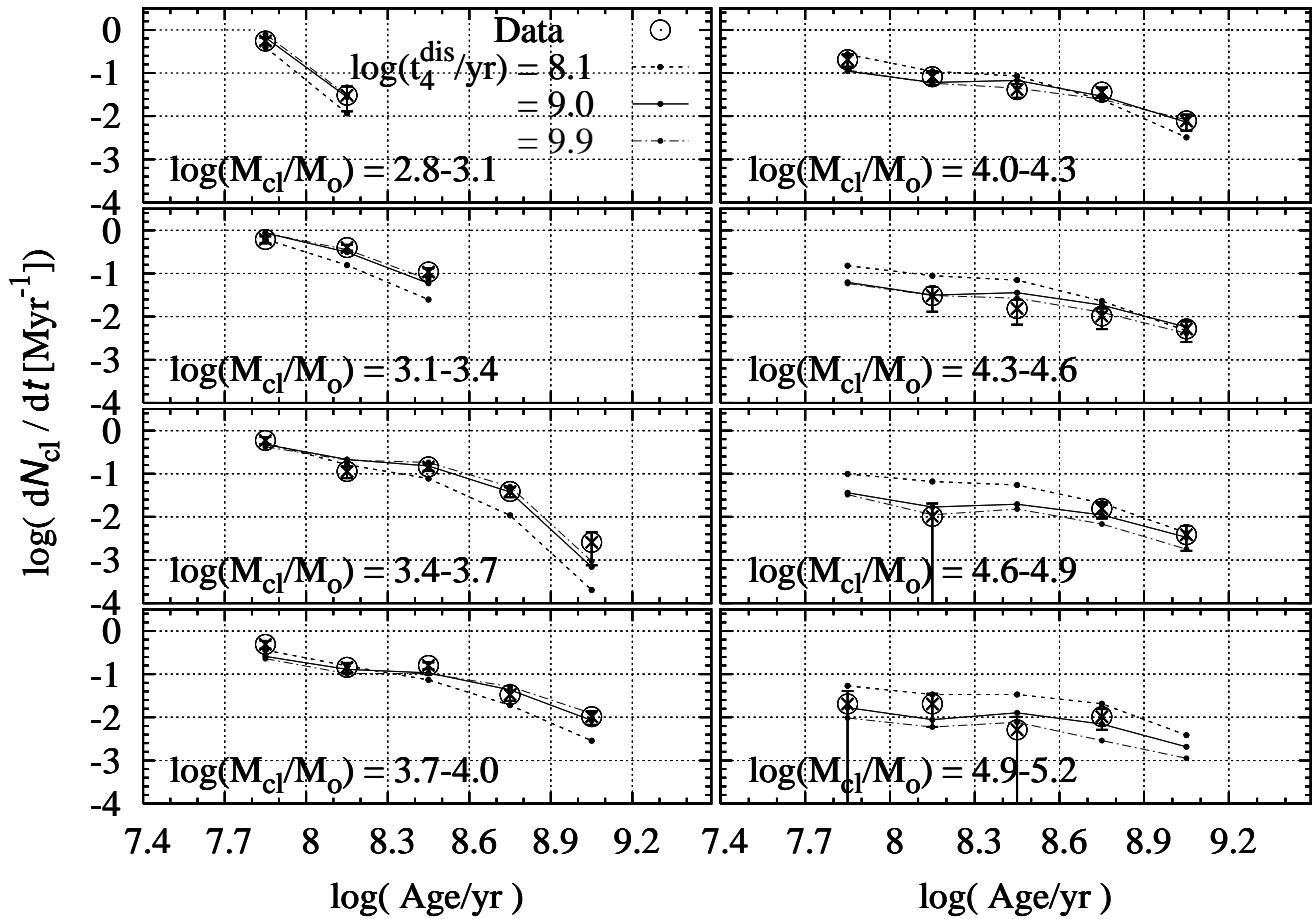, width=\linewidth}
\end{minipage}
\caption{Age distributions of the detected surviving clusters for the
data (large open circles with error bars) and the models (short-dashed
lines: $\log (t_4^{\rm dis} {\rm yr}^{-1})=8.1$; solid lines: $\log
(t_4^{\rm dis} {\rm yr}^{-1})=9.0$; dash-dotted lines:$\log (t_4^{\rm
dis} {\rm yr}^{-1})=9.9$) for the eight mass ranges listed at the
bottom-left of each panel. They correspond to the mass binning used in
Fig.~\ref{fig:obs_am} }
\label{fig:ad_mb} 
\end{figure*}

In order to carefully represent the cluster distribution in the
two-dimensional ($\log({\rm age}),\log(M_{\rm cl})$) space,
Figs.~\ref{fig:MSp_ab} and \ref{fig:ad_mb} show the CMFs per age bin
and the cluster age distributions per mass bin, respectively. Model
and observed distributions are shown by lines and large open circles,
respectively. Note that 1$\sigma$ error bars are often smaller than
the symbol size. The increase with increasing age of the lower mass
limit to the CMF (Fig.~\ref{fig:MSp_ab}) and the age range extension
for greater cluster mass (Fig.\ref{fig:ad_mb}) mirror the detection
limit in ($\log({\rm age}),\log(M_{\rm cl})$) space.

As was concluded based on the top panel of Fig.~\ref{fig:varcfr},
should the LMC be characterised by a short {\dts} of $\log (t_4^{\rm
dis} {\rm yr}^{-1})=8.1$, its CFR would also have been higher by a
factor of $\simeq 5 (10)$ compared to $\log (t_4^{\rm dis} {\rm
yr}^{-1})=9.0 (9.9)$ (see the top panels of Fig.~\ref{fig:cstcfr} and
of Fig.~\ref{fig:varcfr}). Because massive star clusters dissolve on a
longer time-scale than their low-mass counterparts, their age distribution
retains a better imprint of the CFR. The two lower right-hand panels
in Fig.\ref{fig:ad_mb} [$\log(M_{\rm cl}/{\rm M}_{\odot}) > 4.6$] thus
show model age distributions in excess of what is observed if $\log
(t_4^{\rm dis} {\rm yr}^{-1})=8.1$ (short dashed line).  On the other
hand, in the low-mass regime (say, for $M_{\rm cl} < 10^4\,{\rm
M}_{\odot}$), where disruption proceeds on a shorter time-scale, the
synthetic age distributions are under-represented with respect to the
observations (left panels in Fig.~\ref{fig:ad_mb}). Therefore, a
{\dts} of order $\log (t_4^{\rm dis} {\rm yr}^{-1})=8.1$ is ruled out
in the LMC since it would give rise to number counts in the cluster
age distributions that are lower (higher) than the observations for
$\log(M_{\rm cl}/{\rm M}_{\odot}) < 4.0$ ($> 4.6$), {\it despite the
fact that the overall synthetic cluster age distribution,
i.e. integrated over all cluster masses, matches the observations
fairly well}.

In order to complete our analysis, we list in Table
\ref{tab:surv_rate} the ratio $F_M$ between the final total mass and
the initial total mass in clusters, and the corresponding number
count ratio, $F_N$ (i.e. the fraction of surviving clusters), for the three
cluster disruption time-scales considered here and for the case where
we adjusted the CFR to match the observed age distribution. We
consider the age ranges $7.7 < \log(\mbox{age
yr}^{-1}) < 9.2$ (first part of the table) and $7.7 < \log(\mbox{age
yr}^{-1}) < 10.2$ (second part of the table; see section
\ref{sec:agap}). $F_N$ and $F_M$ refer to the survival rates of all
clusters, regardless of their final mass. $F_M^{\rm obs}$ and $F_N^{\rm
obs}$ are the survival rates of the {\it detected} clusters, i.e.,
those brighter than the adopted detection limit at $M_V^{\rm lim} =
-4.7$ mag. As also found in previous studies (Baumgardt 1998;
McLaughlin 1999; Parmentier \& Gilmore 2005), number-related
quantities are more sensitive to dynamical evolution than mass-related
quantities. This is so because, for a power-law ICMF with a spectral
index of $\alpha = -2$, low-mass clusters dominate the initial cluster
population in terms of number, but not in terms of mass, while they
are expected to be disrupted first. For instance, the low-mass range
$2 < \log(M_{\rm cl}/{\rm M}_{\odot}) < 3$ accounts for only 20 per
cent of the total initial mass in clusters (assuming $\alpha = -2$ and
a mass range $10^2 < M_{\rm cl}/{\rm M}_{\odot} < 10^7$), while their
number fraction amounts to about 90 per cent. Finally, we remind the
reader that these survival rates do not take into account infant
mortality.
 
\begin{table}
\begin{center}
\caption[]{Cluster survival rates in terms of number counts and in
terms of mass. $F_N$ is the ratio between the number of surviving
clusters and the initial number of clusters; $F_N^{\rm obs}$ is the
ratio between the number of {\it detected} surviving clusters and the
initial number of clusters. $F_M$ and $F_M^{\rm obs}$ are the
corresponding quantities in units of cluster mass. These figures are
based on a power-law ICMF with a spectral index of $\alpha=-2$, a
detection limit at $M_V^{\rm lim}=-4.7$ mag, and a CFR adjusted so
that the modelled age distribution matches the observed LMC cluster
age distribution. Survival rates are given for three different cluster
disruption time-scales and for two different cluster age ranges.}
\label{tab:surv_rate}
\begin{tabular}{  c c c  } \hline 
      $\log (t_4^{\rm dis} {\rm yr}^{-1})$                          &  ($F_N$, $F_N^{\rm obs}$) &   ($F_M$, $F_M^{\rm obs}$)  \\ \hline
\multicolumn{3}{c}{$ 7.7 < \log(\mbox{age yr}^{-1}) < 9.2 $} \\ \hline
 8.1 & ($4 \times 10^{-3}$,$2 \times 10^{-3}$)   &    (0.20,0.20)     \\ 
 9.0 &   (0.09,0.02)                             &    (0.47,0.44)     \\ 
 9.9 &   (0.49,0.04)                             &    (0.68,0.54)     \\ \hline
\multicolumn{3}{c}{$ 7.7 < \log(\mbox{age yr}^{-1}) < 10.2 $} \\ \hline
 8.1 &  ($5 \times 10^{-4}$,$3 \times 10^{-4}$)  &    (0.03,0.03)       \\ 
 9.0 &        (0.03,0.01)                        &    (0.25,0.23)   \\ 
 9.9 &        (0.32,0.02)                        &    (0.54,0.43)     \\  \hline
\end{tabular}
\end{center}
\end{table}  

\section{Importance of a properly defined detection limit}
\label{sec:comp}

\begin{figure}
\begin{minipage}[t]{\linewidth}
\centering\epsfig{figure=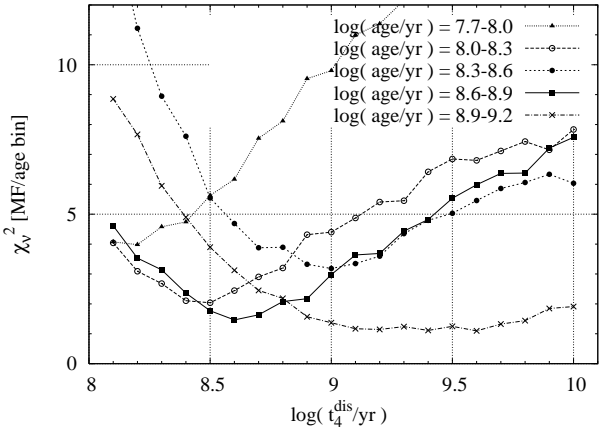, width=\linewidth}
\end{minipage}
\caption{Same as Fig.~\ref{fig:xhi2_MFagebin}, but assuming a fiducial
detection limit decreased by $\Delta \log(M_{\rm cl})=-0.48$ dex
in ($\log ({\rm age}),\log(M_{\rm cl})$) space (i.e., to H03's fading
limit). The minima of the $\chi^2_{\nu}$ functions for the various
CMFs for given age ranges are located at markedly different cluster
disruption time-scales. The inconsistency suggests a problem in the
data, namely, the incorrectly estimated detection limit. That is, CMF
turn-overs created as a result of incompleteness are interpreted in
terms of rapid secular evolution, especially at young ages.}
\label{fig:xhi2_MFagebin_dct} 
\end{figure}

In this section, we investigate how the estimated range of the LMC's
$t_4^{\rm dis}$ responds to vertically shifting the adopted detection
limit in Fig.~\ref{fig:obs_am}. As an extreme case, let us for
instance consider H03's fading limit at $M_V = -3.5$ mag as our
detection limit (dash-dotted line labelled `[3]' in
Fig.~\ref{fig:obs_am}). The turn-over caused by incompleteness
effects in each CMF, for each age bin, would in this case incorrectly
be considered as resulting from dynamical evolution. Assuming a
power-law ICMF of spectral index $\alpha = -2$, the {\dts} can
directly be derived from the turn-over cluster mass for each of the
cluster age ranges considered. The results of this exercise are shown
in Fig.~\ref{fig:xhi2_MFagebin_dct}. The conclusions from the
individual age bins are in conflict with each other. If we consider
the oldest age bin ($8.9 < \log(\mbox{age yr}^{-1}) < 9.2$) only, the
range of acceptable values is $9.0 \lesssim \log (t_4^{\rm dis} {\rm
yr}^{-1}) \lesssim 9.8$. However, in the youngest age bin ($7.7 < \log
(\mbox{age yr}^{-1}) < 8.0$), there is a turn-over at a cluster mass
of about $10^3\,{\rm M}_{\odot}$ (see Fig.~\ref{fig:compl}, although
the age ranges are not strictly identical to those considered in
Fig.~\ref{fig:xhi2_MFagebin_dct}), which would not be expected as
owing to dynamical evolution if the {\dts} derived from the oldest age
range were correct. Consideration of the youngest age bin only
therefore results in a {\dts} estimate of about $\log (t_4^{\rm dis}
{\rm yr}^{-1}) = 8.1-8.2$ (see Fig.~\ref{fig:xhi2_MFagebin_dct}),
i.e. a significantly shorter {\dts} than found for the oldest age
bin. Intermediate values for the {\dts} are derived from consideration
of the three intermediate age bins.

\begin{figure}
\begin{minipage}[t]{\linewidth}
\centering\epsfig{figure=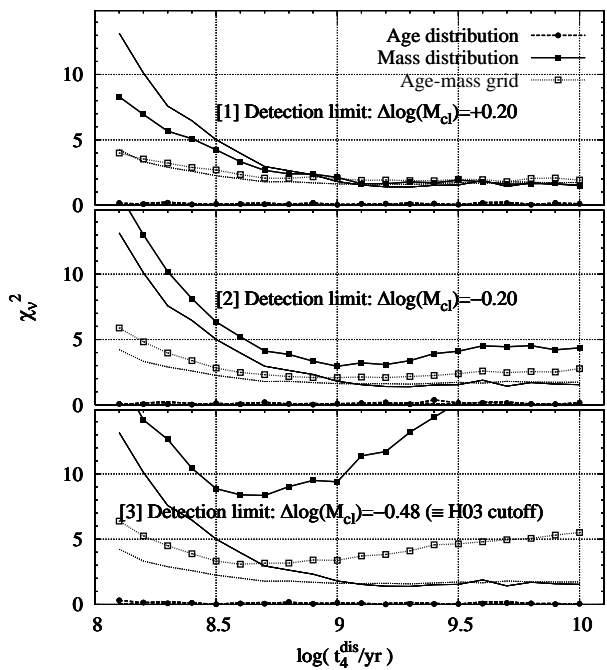, width=\linewidth}
\end{minipage}
\caption{Evolution of $\chi^2_{\nu}$ as a function of {\dts}, assuming
four different detection limits. For comparison,
each panel displays the fiducial case as thin lines. The top, middle
and bottom panels show the results for detection limits
shifted with respect to the fiducial limit by $\Delta \log (M_{\rm
cl})=+0.20, -0.20,$ and $-0.48$, respectively (corresponding to thin
dash-dotted lines labelled `[1]', `[2]'and `[3]' in
Fig.~\ref{fig:obs_am}). The corresponding $\chi^2_{\nu}$ values are
shown as thick lines with symbols.}
\label{fig:rob_xhi2_dct} 
\end{figure}

On a global scale, both the $\chi_{\nu}^2$ obtained from the cluster
mass spectrum integrated over age and the $\chi_{\nu}^2$ from the
distribution of clusters in ($\log({\rm age}),\log(M_{\rm cl})$) space
lead to significantly shorter {\dts} estimates than what we inferred 
from the fiducial case. 
This is shown by the thick curves with filled symbols in
the bottom panel of Fig.~\ref{fig:rob_xhi2_dct}, where we also show
the results for the fiducial case (thin curves). Not only does the
reduced $\chi_{\nu}^2$ derived from the mass spectrum suggest a
shorter $\log (t_4^{\rm dis} {\rm yr}^{-1}) \simeq 8.6-8.7$, it also
increases significantly for longer {\dts}s, thus making those longer
time-scales less probable (see Table \ref{tab:xhi2}). Additionally,
its absolute value is high, $\chi_{\nu ,min}^2 \simeq 8$, indicating a
poor match between the model and the data.

Lacking a robustly established detection limit, one cannot,
therefore, derive a reliable estimate of the {\dts}. There is, in
fact, a degeneracy between the incompleteness effects and the effects
of dynamical evolution, in the sense that the impact of an
underestimated completeness limit is mimicked by a shortened
disruption time-scale.

The top panels of Fig.~\ref{fig:rob_xhi2_dct} show how $\chi^2_{\nu}$
responds to vertical shifts of the fiducial detection limit 
(i.e. $M_V^{\rm lim}=-4.7$) by
$\Delta\log M_{\rm cl} = +0.2$ or $-0.2$ dex (equivalent to $\pm
0.5$\,mag; see the dash-dotted lines labelled `[1]' and `[2]' in
Fig.~\ref{fig:obs_am}). In the first case (top panel), they do not
differ significantly from the fiducial case except for a loss of
sensitivity of the mass spectrum-based $\chi^2_{\nu}$ in the short
{\dts} regime (i.e., these are less significantly rejected). In the
second case (middle panel), the shapes of the $\chi^2_{\nu}$ functions
at $\log (t_4^{\rm dis} {\rm yr}^{-1}) < 9.0$ do not differ from the
fiducial case. However, longer {\dts}s are less probable, which
illustrates the incompleteness-disruption rate degeneracy referred to
above.

\section{The depth of the age gap}
\label{sec:agap}

\begin{figure}
\begin{minipage}[t]{\linewidth}
\centering\epsfig{figure=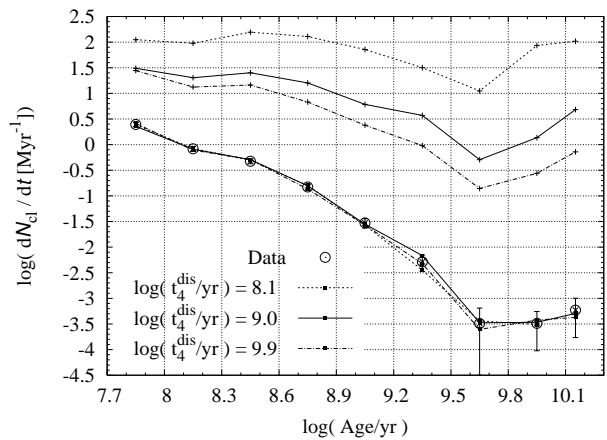, width=\linewidth}
\end{minipage}
\caption{Same as the top panel of Fig.~\ref{fig:varcfr}, except that
the age range has been extended up to 16\,Gyr. Note that this figure
excludes the age range dominated by infant mortality and infant
weight-loss, i.e., the first 50\,Myr. The ICMF is assumed to be a
power law of spectral index $\alpha =-2$, irrespective of when
clusters are formed.}
\label{fig:ad_all} 
\end{figure}

In the previous section, we have put the best possible constraints on
the {\dts} based on the distribution of the detected surviving
clusters in the most populated part of the ($\log({\rm
age}),\log(M_{\rm cl})$) diagram. We now take one step further, and
infer how the LMC's CFR may have varied over the past Hubble
time. Fig.~\ref{fig:ad_all} is the equivalent of the top panel of
Fig.~\ref{fig:varcfr}, but extended up to an age of about 16\,Gyr
(i.e. the maximum cluster age in the {\sc galev} library). As before,
the CFR, adjusted to match the observed age distribution, is shown for
the disruption time-scales, $\log (t_4^{\rm dis} {\rm yr}^{-1}) = 8.1,
9.0,$ and 9.9. The slight scatter in the evolved age distributions
around the data points is entirely caused by sampling noise, because
we limited the number of input clusters to $10^7$ for $\log (t_4^{\rm
dis} {\rm yr}^{-1}) = 8.1$ and to $4 \times 10^5$ for $\log (t_4^{\rm
dis} {\rm yr}^{-1}) = 9.9$. This results in a detected survivor
census of a few $\times 10^3$ clusters.

Let us now consider the extreme values of our best estimate for the
range of cluster disruption time-scales, i.e. $\log (t_4^{\rm dis}
{\rm yr}^{-1})=9.0$ and 9.9. The corresponding evolution of the CFR is
very similar in both cases. From the epoch of the formation of the
first clusters, their number decreases with time, reaching a minimum
about 5\,Gyr ago, before steadily increasing again and remaining
roughly constant for the past 300\,Myr. The present-day formation rate
of the {\it bound} clusters in the LMC is 30 Myr$^{-1}$, regardless of
whether $\log (t_4^{\rm dis} {\rm yr}^{-1})=9.0$ or 9.9. At older
ages, however, both inferred CFRs differ by a factor of 5 (5\,Gyr ago)
up to a factor of 8 ($\simeq 10$\,Gyr ago).  As for the 5\,Gyr-old
``CFR gap'' (mirroring the presently observed cluster age gap, but see
below), this is characterised by a CFR of 30 (for $\log (t_4^{\rm dis}
{\rm yr}^{-1})=9.0$) to 100 (for $\log (t_4^{\rm dis} {\rm
yr}^{-1})=9.9$) times lower than its present-day value.

Previous estimates of the depth of the age gap had already been
provided by H03 and de Grijs \& Anders (2006).  From the temporal
evolution of the maximum cluster mass, and assuming that the mass of
the most massive cluster is determined by size-of-sample effects, H03
infers that the CFR has approximately varied with age as ${\rm CFR}
\propto {\rm age}^{-1}$ between 100\,Myr and 10\,Gyr ago (their
fig.~8). This implies that 5\,Gyr ago, the CFR was $\sim 50$ times
lower than 100\,Myr ago, which matches our own estimates above quite
well. As H03 cautions, however, their result heavily depends on the
age and mass estimates of the few clusters in the age gap.

From the cluster age distribution of de Grijs \& Anders (2006, their
fig.~6a), we derive a seemingly more limited CFR variation.  The
vertical shift between the observed age distribution and the age
distribution predicted for a constant CFR (i.e., their disruption
line) at an age of 5\,Gyr shows an equivalent ratio of $\sim 10$ only.
Since their $\log (t_{\rm cross} {\rm yr}^{-1}) \simeq 8.1$ implies
$t_4^{\rm dis} \simeq 1$\,Gyr, only the comparison to our CFR ratio
for $\log (t_4^{\rm dis}{\rm yr}^{-1})=9.0$ makes sense, that is, a
factor of $\sim 30$ between the CFRs after and inside the
age gap. Careful examination of de Grijs \& Anders' (2006) fig.~6a
reveals that part of the difference is caused by an underestimate of
the slope of their disruption line. While the slope of a constant CFR
disruption line is $(1-\alpha)/\gamma = -1.6$ (Boutloukos \& Lamers 2003), 
their disruption line,
obtained from a simple fit to the data, shows a slope of $\simeq
-1.8$. As we correct for this effect, the ratio of the CFR 100\,Myr
ago to that 5\,Gyr ago is $\sim 20$. In view of the different sampling
limits used by de Grijs \& Anders (2006) and in this paper, these
results (i.e. CFR ratios of 20 and 30) are in reasonable agreement.

As for the CFR at $\simeq$ 10\,Gyr ago, it is worth bearing in mind that
the estimates (3 and 0.5 clusters Myr$^{-1}$ for $\log (t_4^{\rm dis}
{\rm yr}^{-1})=9.0$ and 9.9, respectively) rely thoroughly on the
assumption of a power-law ICMF of $\alpha = -2$ for all clusters,
irrespective of their age.

Whether the first star clusters to form in the Universe (i.e. globular 
clusters) are characterised by that ICMF, typical of young star
clusters in the Galactic disc and in starbursts and mergers
(Zhang \& Fall 1999), or by a Gaussian similar to their
present-day CMF (Vesperini 1998; Parmentier \& Gilmore 2005, 2007)
remains a matter of ongoing debate. It is not our aim in this section
to discuss the true nature of the ICMF in the LMC 13 Gyr ago. We are
prevented from doing so by the paucity of data above the fiducial
detection limit at old age. However, it is worth emphasising
that the fraction by number of clusters that survive to old age
crucially depends on their ICMF. As an illustration, let us assume
that $t_4^{\rm dis}=2.4$\,Gyr. The disruption time-scale of a
$10^5\,{\rm M}_{\odot}$ cluster is then about 10\,Gyr 
(Eq.~\ref{eq:tdis}, assuming $\gamma =0.62$) and all clusters of masses between 
$10^2\,{\rm M}_{\odot}$ and $10^5\,{\rm M}_{\odot}$ will have evaporated by the
time their age reaches a Hubble time. If star clusters formed with an 
ICMF spectral index $\alpha \simeq -2$, this implies a decrease by $\simeq$ 3
orders of magnitude in the initial number of clusters. On the other hand,
if old LMC clusters formed following the universal Gaussian CMF
characteristic of present-day globular cluster systems, clusters more
massive than $10^5\,{\rm M}_{\odot}$ account for about one half of the
initial cluster census, thereby implying a survival rate of $\simeq$
50 per cent.  Also, the cluster survival rates in terms of mass
and in terms of number are similar in that case (see table 3 in 
Parmentier \& Gilmore~2005 for a comparison of the Galactic halo 
cluster survival rates in case of power-law and Gaussian ICMFs).
Since the survival rates of clusters (in terms of number) at old age 
depend {\it heavily} on their ICMF, so does the inferred CFR.

\begin{figure}
\centering\epsfig{figure=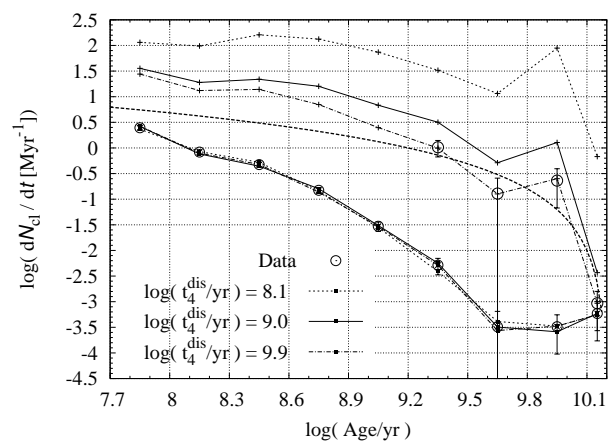, width=\linewidth}
\caption{Same as Fig.~\ref{fig:ad_all}, except for the choice of the
ICMF for the oldest age bin: $10.1 < \log({\rm age} {\rm  yr}^{-1}) < 10.2$, for
which we assume the universal Gaussian globular cluster mass function. 
Note how different the predicted CFR at old age is
compared to Fig.~\ref{fig:ad_all}. With an ICMF similar to the
universal Gaussian globular cluster mass function, only a few low-mass
(say less massive than $10^4\,{\rm M}_{\odot}$) clusters form, which
boosts the cluster survival rate and correspondingly lowers the 
CFR at old age. Although the temporal evolution of the CFR is
rather noisy at old and intermediate ages, it remains consistent (at
the $1\sigma$ level) with a smooth and steady increase, as shown by
the thick dashed line.}
\label{fig:ad_oldG} 
\end{figure}

In Fig.~\ref{fig:ad_oldG}, we consider the universal globular cluster
mass function for the oldest age bin. In other words, in the age range
$10.1 < \log (\mbox{age yr}^{-1}) < 10.2$, we substitute the power-law
ICMF of slope $\alpha = -2$ assumed thus far by a Gaussian ICMF with a
mean logarithmic cluster mass of 5.3 and a standard deviation of 0.6
dex. At first glance, the evolution of the CFR looks rather noisy. The
very low CFR at old age (3 and 1 clusters Gyr$^{-1}$ for $\log
(t_4^{\rm dis} {\rm yr}^{-1})=9.0$ and 9.9, respectively) is followed
by a steep rise amounting to a factor of order 300, which is
subsequently succeeded by a 3-fold decrease leading to the age gap.

However, because the CFR is directly determined from the present-day
age distribution, they are both affected by the same uncertainties. At
ages older than 3\,Gyr, the $1\sigma$ error bars of the age
distribution become significantly larger than at younger ages, as a
result of the smaller number of clusters. The CFR at intermediate and
old ages becomes similarly uncertain. In order to highlight this, in
Fig.~\ref{fig:ad_oldG} we plot the Poissonian error bars of the four
oldest bins of the age distribution on the CFR for $\log (t_4^{\rm
dis} {\rm yr}^{-1})=9.9$. From this simple exercice, it appears that
the CFR may have steadily increased from a very low value of about 1
to 3 clusters per Gyr some 15\,Gyr ago, up to a $300-1000$ times
higher rate 2 Gyr ago. This steady increase in the CFR is indicated by
the thick dashed line in Fig.\ref{fig:ad_oldG}; this is compatible
with the four oldest age bins within $1\sigma$.

We therefore conclude that the well-established age gap in the LMC
cluster age distribution may not be a faithful mirror of the
underlying CFR. This is mainly owing to the assumption that all star
clusters were formed in a power-law fashion. For this scenario, the
CFR develops a minimum at about 5\,Gyr ago, and increases towards both
older and younger ages around the age gap (see Fig.~\ref{fig:ad_all}). 
However, if the ICMF at the oldest age is
different, and similar to the observed Gaussian globular cluster mass
function, the resulting bias towards greater cluster mass boosts the
number fraction of surviving clusters, thus resulting in a much lower
CFR at old age. In the latter case, the CFR has steadily increased
over the past Hubble time (see Fig.~\ref{fig:ad_oldG}). 
  
Although, in that model, the CFR is increasing by about 2 orders of magnitude 
from 16\,Gyr ago to 8\,Gyr ago, the corresponding variations 
of the formation rate in terms of cluster {\it mass} are not that significant.  
Referring to the thick dashed curve in Fig.\ref{fig:ad_oldG} (i.e. that 
mimicing the smooth and steady CFR increase with time), the CFR in the 
last and last but one age bins are $\simeq 1$ cluster per Gyr and 
$\simeq 50$ clusters per Gyr, respectively.  Owing to a different 
underlying ICMF, these two subsequent age bins are also characterized by markedly
different initial mean cluster mass, $\simeq 1000\,{\rm M}_{\odot}$ for an
$\alpha =-2$ power-law ICMF and $\simeq 10^5\,{\rm M}_{\odot}$ for the 
universal Gaussian ICMF.  Therefore, even though at an age of 10\,Gyr the CFR 
is increasing sharply, the total mass formed in clusters per unit of time 
is remaining roughly constant, that is, of order 
$10^2\,{\rm M}_{\odot}.{\rm Myr}^{-1}$. 

For the sake of completeness, Table \ref{tab:initial} lists the initial number 
of clusters $N_{\rm init}$ and the initial total mass in clusters $M_{\rm init}$ 
required to account for the census of LMC clusters with 
$ 7.7 < \log(\mbox{age yr}^{-1}) < 10.2 $.  
If the ICMF adopted for the oldest age bin is the observed Gaussian globular
cluster mass function (quoted by "PL+G$_{\rm old}$" in Table \ref{tab:initial}), 
the increased cluster survival rate at old age lowers the initial size and the 
inital total mass of the cluster population, although the effect remains limited 
owing to the corresponding narrow age range.  

While Table \ref{tab:initial} builds on a cluster dissolution model with 
$\gamma =0.62$, we have also run simulations with $\gamma =1$ 
(see section \ref{sec:disr} and equations \ref{cl_mass_evol_L.eq} and~\ref{eq:tdis}).  
The size and total mass of the initial cluster population are very much the
same if $\log (t_4^{\rm dis}{\rm yr}^{-1})=9.9$.  By virtue of the weakness of 
the disruptive processes, the model shows little sensitivity to $\gamma$ variations.
In case of $\log (t_4^{\rm dis}{\rm yr}^{-1})=9.0$, the variation in the initial
cluster census does not exceed 25 per cent of what we found for $\gamma =0.62$.   \\
 
\begin{table}
\begin{center}
\caption[]{Initial number of clusters $N_{\rm init}$ and initial total mass in 
clusters $M_{\rm init}$ required to account for the population of LMC clusters with 
$ 7.7 < \log(\mbox{age yr}^{-1}) < 10.2 $.  Values are provided for two acceptable 
cluster disruption time-scales, as well as for two distinct
ICMFs for the oldest clusters, either the canonical power-law ("PL") or the universal
Gaussian globular cluster mass function ("PL+G$_{\rm old}$")}
\label{tab:initial}
\begin{tabular}{  c c c c } \hline 
 ICMF             &  $\log (t_4^{\rm dis}{\rm yr}^{-1})$  &           $M_{\rm init}$             &  $N_{\rm init}$    \\ \hline
 PL               &                       9.0             &   $60 \times 10^6\,{\rm M}_{\odot}$   &  $52 \times 10^3$  \\ 
 PL               &                       9.9             &   $18 \times 10^6\,{\rm M}_{\odot}$   &  $16 \times 10^3$  \\  \hline
 PL+G$_{\rm old}$ &                       9.0             &   $46 \times 10^6\,{\rm M}_{\odot}$   &  $34 \times 10^3$  \\ 
 PL+G$_{\rm old}$ &                       9.9             &   $17 \times 10^6\,{\rm M}_{\odot}$   &  $14 \times 10^3$  \\  \hline
\end{tabular}
\end{center}
\end{table}  

The cluster formation history of the LMC reflects its interaction
history with the Small Magellanic Cloud (SMC) and with the Galaxy.
This, in turn, is strongly influenced by the orbital evolution of the
Magellanic Clouds, which remains poorly known, however. It is still
unclear whether the Magellanic Clouds were born initially as a
primordial pair of galaxies, or whether they dynamically coupled for
the first time some 4\,Gyr ago. Besla et al.~(2007) re-examine the
orbital history of the Clouds on the basis of the most recent proper
motion measurements available (Kallivayalil et al.~2006a,b). They find
that the existence of a stable binary LMC-SMC system lasting for a
Hubble time cannot be precluded, although that remains uncertain. The
present status of the orbital history of the Magellanic Clouds about
the Galaxy is equally questionable.  According to most previous
studies, the LMC has completed several pericentric passages about the
Galaxy in a Hubble time. Besla et al.~(2007) reached the opposite
conclusion, however, that the LMC-SMC pair of galaxies may be on their
first passage about the Milky Way.

The orbital history of the LMC remains therefore poorly
constrained. And so are the origins of the temporal variations of its
CFR. Why did it increase from 3\,Gyr ago? Whereas Bekki \& Chiba
(2005) advocate violent and close interactions with the SMC, Besla et
al.~(2007) favour a scenario in which interactions with the Galaxy are
the dominant factor. In fact, the CFR increase from $\sim 3$\,Gyr ago
is a natural consequence of their first-passage scenario, as this
corresponds to when the Magellanic Clouds first enter inside the
virial radius of the Milky Way and begin to interact with the Galactic
halo gas.

Finally, we are left with the question as to whether cluster formation
in the LMC was in fact reactivated some 3 to 4 Gyr ago (i.e., the ICMF
at old age is a power law of spectral index $\alpha =-2$) or whether
this increase is merely a continuation of a process started $\sim
15$\,Gyr ago (i.e., the ICMF at old ages is the universal Gaussian
globular cluster mass function). A decrease in the CFR from 15\,Gyr
ago until $\sim 5$\,Gyr ago may be understood in the context of a
scenario in which the evolution of the LMC-SMC-Milky Way system is not
considered in isolation but, instead, in the wider context of the
evolution of the entire Local Group of galaxies. Shuter (1992) and
Byrd et al.~(1994) considered the possibility that the Magellanic
Clouds left the neighbourhood of the Andromeda Galaxy about 10 Gyr ago
and were only recently tidally captured by the Milky Way.  While
recent tidal capture by the Milky Way could have reactivated cluster
formation, the decreasing cluster formation rate from the earliest
stages may then result from the LMC moving to greater distances from
M31. Yet, this scenario is not supported by Besla et al.'s (2007)
computations. On the other hand, the steady increase in
Fig.~\ref{fig:ad_oldG} may result from the LMC getting closer to the
Milky Way, in the context of the first perigalactic passage scenario
of Besla et al.~(2007). Only future astrometric missions reaching
$\sim 10 \mu$arcsec accuracy can help clarify these issues.

\section{Disentangling the 1 and 10\,Gyr cluster disruption time-scales: what kind of data do we miss ?}
\label{sec:1vs10}

\begin{figure}
\begin{minipage}[t]{\linewidth}
\centering\epsfig{figure=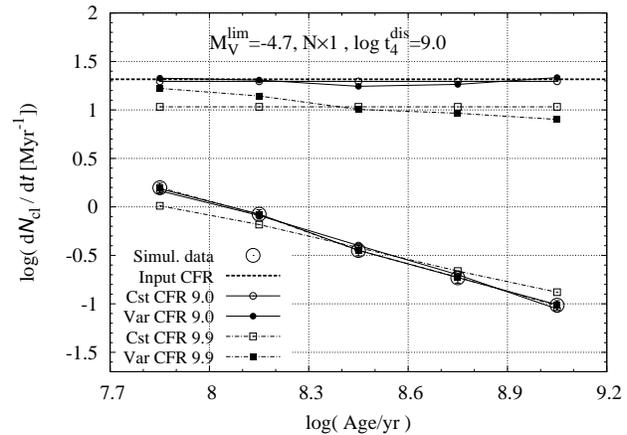, width=\linewidth}
\end{minipage}
\caption{Observed simulations. Simulated data (large open circles)
consist of a putative population of star clusters formed with an
$\alpha =-2$ power-law ICMF and a constant CFR (indicated by the
horizontal thick dashed line) over the age range of the figure. We
adopt $t_4^{\rm dis}=1$\,Gyr and the fiducial detection limit
of Fig.~\ref{fig:obs_am}. The upper lines show the temporal variations
of the CFR as derived from the analysis done with the simulated
data. Open symbols represent a constant CFR, while filled symbols show
CFRs adjusted to match the evolved and observed age
distributions. Solid (dash-dotted) lines represent simulations with
$\log (t_4^{\rm dis} {\rm yr}^{-1})=9.0$ (9.9). Lower lines,
identically coded, are the corresponding age distributions. }
\label{fig:simul_oad} 
\end{figure}

In Section \ref{sec:disr}, we estimated the characteristic disruption
time-scale $t_4^{\rm dis}$ of a $10^4\,{\rm M}_{\odot}$ cluster in the 
LMC. We concluded that the best constraint we could obtain is a 
{\it range} of estimates, i.e. $t_4^{\rm dis} \ge 1$\,Gyr. In other 
words, despite the proximity of the LMC to the Milky Way, its {\dts} 
cannot be constrained to better than an order of magnitude. 
 
In this section, we investigate which strategy we may adopt in the
future to tighten the constraint to the {\dts} and, therefore, the
cluster formation history in the LMC. Specifically, we generate sets
of ($\log({\rm age}),\log(M_{\rm cl})$) data, which we subsequently
``observe'' following the same procedure as before. We can then
compare the retrieved (range of) {\dts} estimates with the disruption
rate actually used to generate these synthetic cluster populations.

\begin{figure*}
\centering\epsfig{figure=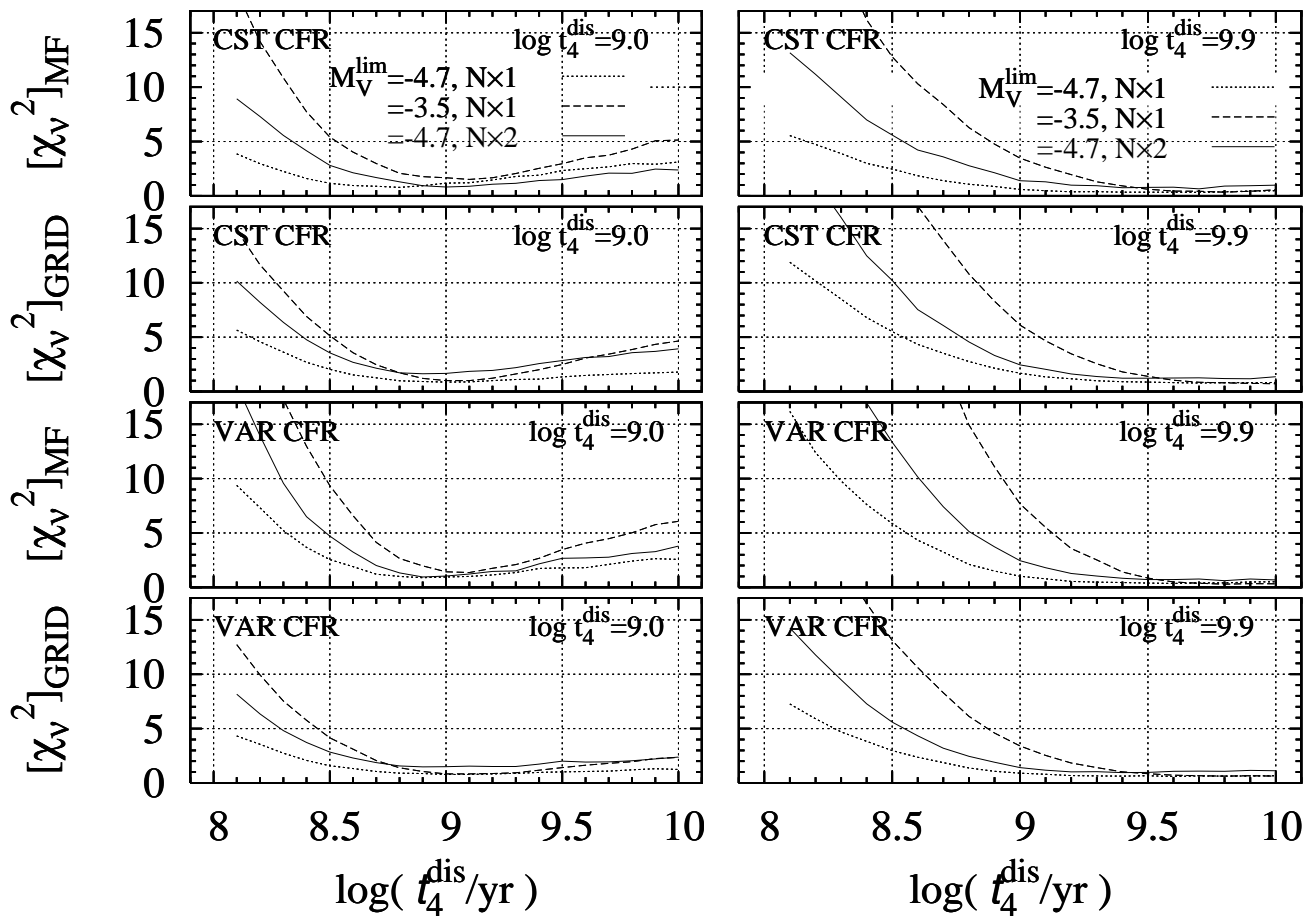, width=\linewidth}
\caption{Results of $\chi^2$ tests done for synthetic star cluster
populations. These are formed with a constant CFR, the ``canonical''
power-law ICMF ($\alpha =-2$), and cluster disruption time-scales of
$\log (t_4^{\rm dis} {\rm yr}^{-1})=9.0$ and 9.9 in the left and
right-hand panels, respectively. Following the addition of Gaussian
noise to $\log(M_{\rm cl})$ and $\log({\rm age})$ of the survivors (to
mimic an observational situation), clusters with $7.7 < \log(\mbox{age
yr}^{-1}) < 9.2$ and located above a detection limit of either
$M_V^{\rm lim}=-4.7$ mag (solid and dotted lines) or $M_V^{\rm
lim}=-3.5$ mag (dashed line) are extracted from the generated files
and processed by the same method as used in Section \ref{sec:disr} for
the LMC data. The resultant $\chi^2_{\nu}$ functions are shown. Dotted
lines correspond to a star cluster system including about 380 clusters
brighter than $M_V^{\rm lim}=-4.7$ mag (cf. Section
\ref{sec:disr}). Solid lines illustrate how $\chi^2_{\nu}$ functions
improve should the cluster census be doubled, all other parameters
being kept the same. The dashed lines show the accuracy of the {\dts}
that is to be expected if the LMC cluster sample were complete above
$M_V^{\rm lim}=-3.5$ mag.}
\label{fig:sim_xhi2} 
\end{figure*}

Six distinct cases will be discussed. We consider two cluster
disruption time-scales ($\log (t_4^{\rm dis} {\rm yr}^{-1})=9.0$ and
9.9), combined with three distinct observational situations, in terms
of cluster number and detection limit. First, we assume a detection
limit and a number of observed clusters similar to that in Section
\ref{sec:disr} (i.e., $\simeq 400$ clusters brighter than $M_V^{\rm
lim}=-4.7$ mag and spanning the age range [50\,Myr,1.5\,Gyr]; labelled
"$M_V^{\rm lim}=-4.7$, $N \times 1$" in Figs.~\ref{fig:simul_oad} and 
\ref{fig:sim_xhi2}). Secondly, we retain the same detection limit but
consider twice as many clusters (i.e., $\simeq 800$ clusters brighter
than $M_V^{\rm lim}=-4.7$ mag, for $7.7 < \log(\mbox{age yr}^{-1}) <
9.2$; labelled "$M_V^{\rm lim}=-4.7$, $N \times 2$" in Fig.~\ref{fig:sim_xhi2}). 
And thirdly, we consider the same number of clusters brighter
than $M_V^{\rm lim}=-4.7$ mag, but a detection limit improved by 1.2\,mag,
at $M_V^{l\rm im}=-3.5$ (the fading limit of H03 represented by the dash-dotted 
line '[3]' in Fig.~\ref{fig:obs_am}; labelled
"$M_V^{\rm lim}=-3.5$, $N \times 1$" in Fig.~\ref{fig:sim_xhi2}). 
In the latter case, the total number of clusters is significantly 
larger because of the inclusion of objects with $-4.7 < M_V < -3.5$ mag.
 
For the sake of simplicity, we have assumed a constant CFR for $7.4 <
\log(\mbox{age yr}^{-1}) < 9.5$ and the ``canonical'' $\alpha=-2$
ICMF. After generating ages and masses for the required number of
clusters, we added Gaussian noise to our synthetic data set. We adopt
$\sigma = 0.15$ dex as the standard deviation of logarithmic cluster
ages and masses.  This represents a fair upper limit to the internal
errors for the de Grijs \& Anders (2006) sample, since $\simeq 80$\,
per cent of their clusters exhibit smaller errors. We then extracted
clusters in the appropriate age range and above the appropriate
detection limit. Finally, their age distribution integrated over mass,
mass distribution integrated over age, and the two-dimensional ($\log({\rm
age}),\log(M_{\rm cl})$) distribution were analysed in exactly the
same way as we did for the observed LMC cluster sample (Section
\ref{sec:disr}).

\begin{figure*}
\centering\epsfig{figure=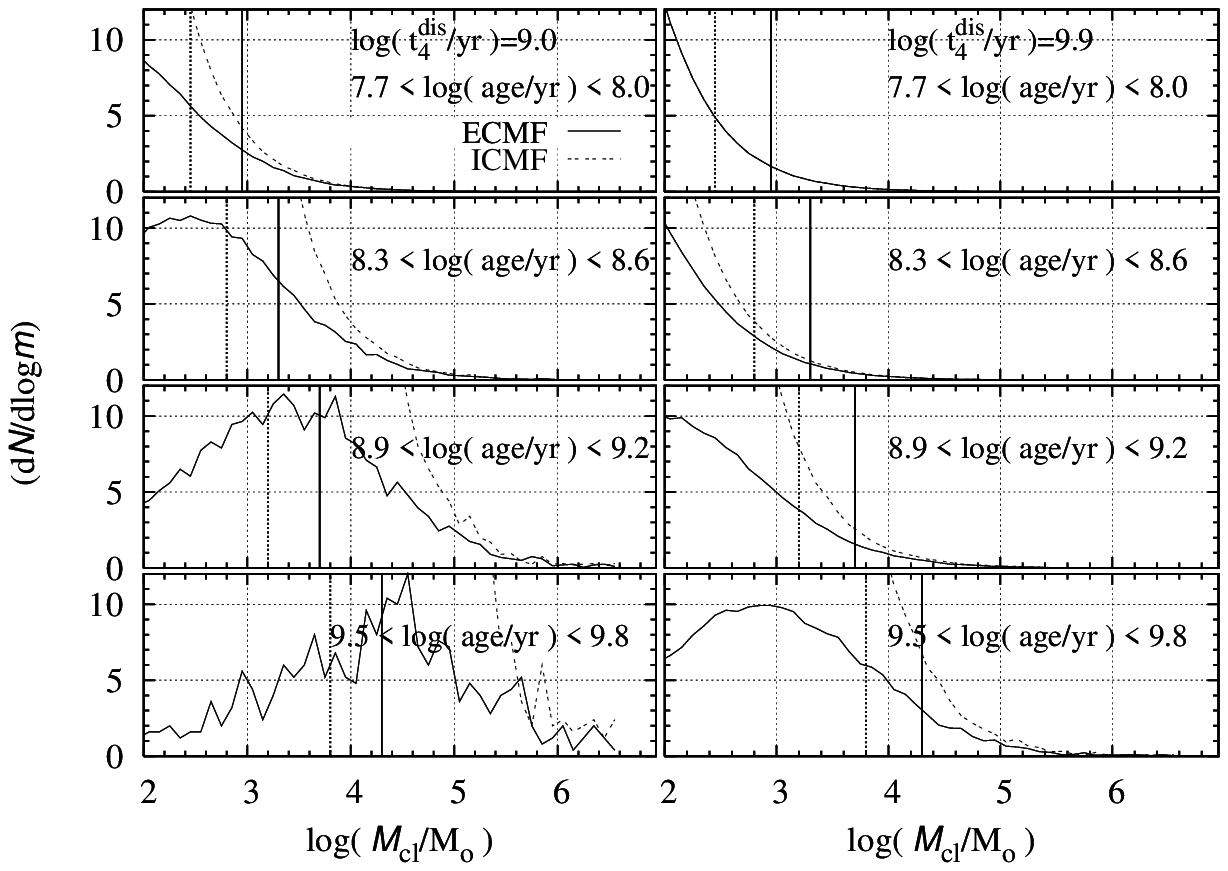, width=\linewidth}
\caption{Evolved CMFs (solid lines) of synthetic star cluster
populations formed with a constant CFR, a power-law ICMF of spectral
index $\alpha =-2$ (dashed lines), and two cluster disruption
time-scales of $\log (t_4^{\rm dis} {\rm yr}^{-1})=9.0$ and 9.9 (left
and right-hand panels, respectively). Different age ranges (listed in
each panel) are displayed. The vertical solid and dashed lines
represent detection limits at $M_V^{\rm lim}=-4.7$ mag (fiducial
detection limit; thick dash-dotted line with symbols in Fig.~\ref{fig:obs_am}) 
and $M_V^{\rm lim}=-3.5$ mag (thin dash-dotted line labelled `[3]' in
Fig.~\ref{fig:obs_am}). As secular evolution proceeds, the turn-over
in mass of the dynamically-shaped CMF shifts towards higher cluster
masses and the number of surviving clusters steadily declines, thereby
enhancing the presence of sampling noise in the CMF. The vertical
scaling is arbitrary.}
\label{fig:tevol_MSp} 
\end{figure*}

Figure \ref{fig:simul_oad} illustrates the simulated observed age
distribution (large open circles) obtained with $\simeq 400$ surviving 
clusters brighter than the fiducial detection limit at $M_V^{lim}=-4.7$
in case of $\log (t_4^{\rm dis} {\rm yr}^{-1})=9.0$. The upper four lines are the derived CFRs, either
constant (open symbols) or adjusted to match the model and observed
age distributions (filled symbols). The lower lines, using the same line
style coding, are the model age distributions of the detected
survivors. Model solutions are shown for two disruption time-scales,
$\log (t_4^{\rm dis} {\rm yr}^{-1})=9.0$ (circle-symbol lines) and 9.9
(square-symbol lines). 

The observed age distribution is well-matched for a constant CFR
combined with a cluster disruption time-scale of $\log (t_4^{\rm dis}
{\rm yr}^{-1})=9.0$, since these are the actual characteristics of the
simulated observations. On the other hand, the model based on a
constant CFR and $\log (t_4^{\rm dis} {\rm yr}^{-1})=9.9$ results in
an age distribution (dash-dotted line with open squares) that is
shallower than the ``observations''.  Therefore, on average, the
corresponding {\it adjusted} CFR declines towards older age
(dash-dotted line with filled squares).

The input constant CFR used to generate the ``observations'' is shown
as the horizontal thick dashed line (i.e., 20 clusters Myr$^{-1}$). One
can immediately see that, if the {\dts} is correctly estimated, the
CFR can be derived with great accuracy (i.e., the thick dashed line
and the solid line with circles follow each other closely), except for
some sampling noise.  If the {\dts} is
incorrectly assumed to be 8 times longer ($\log (t_4^{\rm dis} {\rm
yr}^{-1})=9.9$), then the CFR is underestimated (as a result of the
assumed slow cluster disruption), although by an average factor of two
at most.

We now investigate what the conditions are that we need to derive a 
better approximation to the LMC cluster formation history than what we
found in Section \ref{sec:disr}.

In Fig.~\ref{fig:sim_xhi2} we plot the evolution of the ($t_4^{\rm
dis}$, $\chi^2_{\nu}$) functions which follow from the analysis of the
simulated data. In the left and right-hand panels, the simulated star
cluster evolution is characterised by $\log (t_4^{\rm dis}{\rm
yr}^{-1})=9.0$ and 9.9, respectively. $[\chi^2_{\nu}]_{\rm MF}$ and
$[\chi^2_{\nu}]_{\rm GRID}$ are the reduced $\chi^2_{\nu}$ derived
from the CMF integrated over age, and from the two-dimensional
($\log({\rm age}),\log(M_{\rm cl})$) distribution of clusters,
respectively. 

For an observational situation similar to that discussed in this paper
(dotted line), we see that the {\dts} is rather loosely constrained.
In none of the panels does the ($t_4^{\rm dis}$, $\chi^2_{\nu}$)
function show a clear minimum. For instance, let us consider the third
panel from the top in the left column of Fig.~\ref{fig:sim_xhi2}
(i.e., the $[\chi^2_{\nu}]_{\rm MF}$ obtained for an adjusted CFR --
`Var CFR'). It shows that $[\chi^2_{\nu}]_{\rm MF, min} \simeq 1$ at
$\log (t_4^{\rm dis}{\rm yr}^{-1})=8.9$ and $[\chi^2_{\nu}]_{\rm MF} <
[\chi^2_{\nu}]_{\rm MF,min}+1$ for $8.6 \leq \log (t_4^{\rm dis}{\rm
yr}^{-1}) \leq 9.7$. We therefore recover the correct {\dts} ($\log
(t_4^{\rm dis}{\rm yr}^{-1})=9.0$), but the range of acceptable
estimates covers a decade in $t_4^{\rm dis}$. Even if the detection
limit remains the same, that situation is improved if the number of
clusters in the sample is increased. With twice as many clusters, one
now obtains $[\chi^2_{\nu}]_{\rm MF, min} \simeq 1$ at $\log
(t_4^{\rm dis}{\rm yr}^{-1})=8.9$ and $[\chi^2_{\nu}]_{\rm MF} <
[\chi^2_{\nu}]_{\rm MF,min}+1$ for $8.7 \leq \log (t_4^{\rm dis}{\rm
yr}^{-1}) \leq 9.4$ (solid curve). The uncertainties on $\log
(t_4^{\rm dis})$ are now reduced by a factor of about two, as a direct
result of the smaller Poissonian noise in the cluster age and mass
distributions. In other words, the more populous a star cluster system
is, the better its inferred cluster formation history.

The best accuracy for the $t_4^{\rm dis}$ estimate is achieved when
the detection limit gets fainter by 1.2\,mag (dashed curve). In that case,
$[\chi^2_{\nu}]_{\rm MF, min} \simeq 1.4$ at $\log (t_4^{\rm dis}{\rm
yr}^{-1})=9.1$ and $[\chi^2_{\nu}]_{\rm MF} < [\chi^2_{\nu}]_{\rm
MF,min}+1$ for $8.8 \leq \log (t_4^{\rm dis}{\rm yr}^{-1}) \leq 9.35$. 
If we compare this with the right-hand
equivalent panel in Fig.~\ref{fig:sim_xhi2}, for which $\log (t_4^{\rm
dis}{\rm yr}^{-1})=9.9$, we find $[\chi^2_{\nu}]_{\rm MF, min} \simeq
0.3$ for $\log (t_4^{\rm dis}{\rm yr}^{-1}) \geq 9.7$ and
$[\chi^2_{\nu}]_{\rm MF} < [\chi^2_{\nu}]_{\rm MF,min}+1$ for $\log
(t_4^{\rm dis}{\rm yr}^{-1}) \geq 9.4$. We therefore conclude that if
the star cluster system of the LMC were observed with a 90 per cent
completeness limit of $M_V \simeq -3.5$ mag, we could decide more
easily whether the {\dts} is of order 1 or 10\,Gyr, which we cannot do
with the data presently available to us.

The reason why a detection limit improved by $\Delta \log(M_{\rm
cl}/{\rm M}_\odot) \simeq -0.5$ better distinguishes $\log (t_4^{\rm
dis} {\rm yr}^{-1})=9.0$ from 9.9 is directly related to whether or
not the turn-over expected to appear in the CMF as a result of
dynamical secular evolution shows up above the detection limit. To
illustrate this, we carried out additional simulations, the results of
which we show in Fig.~\ref{fig:tevol_MSp}.

Each panel of Fig.~\ref{fig:tevol_MSp} corresponds to a synthetic star
cluster population formed with a constant CFR for the given age range,
and with the ``canonical'' ($\alpha =-2$) power-law ICMF. Here, we
are interested in the location of the CMF turn-over with respect to
the detection limit, independently of the presence of observational
errors and of size-of-sample related Poissonian noise.  To limit this,
we adopt a fairly large initial number of clusters ($4 \times 10^5$)
for each limited age range (i.e., for each panel).  The {\dts} is
$\log (t_4^{\rm dis} {\rm yr}^{-1})=9.0$ and 9.9 in the left and
right-hand panels, respectively. The dashed line represents the ICMF
and the solid line is for the evolved CMF for the age ranges
considered, i.e., from top to bottom: $7.7 < \log(\mbox{age yr}^{-1})
< 8.0$, $8.3 < \log(\mbox{age yr}^{-1}) < 8.6$, $8.9 < \log(\mbox{age
yr}^{-1}) < 9.2$, and $9.5 < \log (\mbox{age yr}^{-1}) < 9.8$. Also
shown are the detection limits corresponding to
$M_V^{\rm lim} \simeq -4.7$ mag (vertical thick solid line) and to $M_V^{\rm
lim} \simeq -3.5$ mag (vertical thick dotted line), which we discussed above.

At young ages ($7.7 < \log(\mbox{age yr}^{-1}) < 8.0$), the evolved
CMF remains a good proxy to the ICMF. As clusters age, the power-law
ICMF is secularly turned into a Gaussian CMF. The older the cluster
system and/or the shorter $t^{\rm dis}_4$, the higher the resulting
cluster mass at the turn-over. Let us consider the left-hand panels,
for which $\log (t_4^{\rm dis} {\rm yr}^{-1})=9.0$. At an age of
1\,Gyr, the turn-over remains at the lower edge of the fiducial
detection limit of $M_V^{\rm lim} \simeq -4.7$ mag. As a result, the
imprint of dynamical evolution is barely detectable and $t_4^{\rm
dis}$ cannot be determined from the turn-over location. All we can
retrieve is a lower limit to $t_4^{\rm dis}$, in the sense that we can
exclude all disruption time-scales which are short enough to bring the
CMF turn-over beyond the detection limit. With the improved detection
limit of $M_V^{\rm lim} \simeq -3.5$ mag (vertical dotted line), the
cluster mass at the turn-over is (just) measurable, hence the clear minimum
seen in the $[\chi^2_{\nu}]_{\rm MF}$ of the third panel on the
left-hand side in Fig.~\ref{fig:sim_xhi2} (dashed line). If $\log
(t_4^{\rm dis} {\rm yr}^{-1})=9.9$, however, the CMF evolves on such a
slow time-scale that the turn-over remains undetected, regardless of
whether $M_V^{\rm lim} \simeq -3.5$ or $-4.7$ mag. Only a lower limit
to $\log (t_4^{\rm dis} {\rm yr}^{-1})$ can be inferred and the
$\chi^2_{\nu}$ functions are steadily decreasing with increasing
$t_4^{\rm dis}$ (right-hand panels of Fig.~\ref{fig:sim_xhi2}).

We note in passing that deriving a $t_4^{\rm dis}$ estimate rather
than a lower limit is better achievable based on a more evolved
cluster population, if either the {\dts} is shorter or if the
population contains older clusters. As for the latter, this is so
because disruption limits in ($\log({\rm age}),\log(M_{\rm cl})$)
space (dashed lines in Fig.~\ref{fig:obs_am}) are steeper than
detection limits (dash-dotted lines in Fig.~\ref{fig:obs_am}). Yet, in
that case, the small number of survivors at old age, where the
turn-over location is firmly set, may hamper any analysis as a result
of significant Poissonian noise (see, e.g., the fourth panel on the
left-hand side in Fig.~\ref{fig:tevol_MSp}).

Our LMC star cluster sample lacks clusters at intermediate and old
age, as it contains only $\simeq$ 20 star clusters older than 1\,Gyr
above the fiducial detection limit. Combined with the bright fiducial
detection limit we defined in Section \ref{sec:synth}, this
data paucity results in only a lower limit to the characteristic LMC
star cluster disruption time-scale of $t_4^{\rm dis} \geq 1$\,Gyr.

\section{Summary and Conclusions}
\label{sec:conc}

In this paper, we have carried out numerous detailed Monte-Carlo
simulations aimed at constraining the cluster formation history and
the rate of bound cluster disruption in the LMC star cluster
system. We considered only clusters older than 50\,Myr in order not to
determine erroneously short cluster disruption time-scale as a result
of the inclusion of the infant mortality and infant weight loss
evolutionary phases. Our data are based on the $UBVR$ photometry of H03
(obtained from Massey's (2002) survey of the Magellanic Clouds), for
which de Grijs \& Anders (2006) obtained homogeneously determined age
and mass estimates. We estimated a fiducial detection limit above which
the cluster sample is (fairly) complete from the CMF as a function of 
age, at $M_V^{\rm lim} = -4.7 {\rm mag}$.  This is significantly 
brighter than H03's fading limit. We consider only
clusters brighter than this limit, in order
to avoid severe statistical incompleteness effects.

We have evolved synthetic star cluster systems characterised by
constant CFRs assuming 20 different {\dts}s. The CFR was adjusted to
reproduce the observed age distribution. By doing so, we are likely to
loose sensitivity to CFR variations occurring on a time-scale shorter
than the width of the age range corresponding to each logarithmic age
bin (i.e., more likely at old age where our age bins span greater
linear age ranges than at young age). However, the general behaviour
of the CFR, as well as the CFR averaged over age are robustly
recovered (see Section \ref{sec:1vs10}). We then compare, in a
``Poissonian'' $\chi^2$ sense the modelled mass distribution and the
modelled [$\log({\rm age}),\log(M_{\rm cl})$] distribution to the
observations. We show that because of the bright detection limit at
$M_V^{\rm lim}= -4.7$ mag, one cannot constrain $t_4^{\rm dis}$ to
better than a lower limit, $t_4^{\rm dis} \ge 1$\,Gyr. The tightest
constraints are set by the CMF integrated over age. The $\chi^2$ test
applied to the distribution of points in [$\log({\rm age}),\log(M_{\rm
cl})$] space turns out to be a poor diagnostic tool. This is probably
related to the low density of points in each cell of the [$\log({\rm
age}),\log(M_{\rm cl})$], compared to the density of points in each bin
of the integrated CMF, leading to smaller Poissonnian error bars and
therefore a better constrained model in the latter case. Our range of
{\dts} estimates is robust with respect to model variations, such as
of the upper limit of the initial cluster mass range, the location of
the grid in [$\log({\rm age}),\log(M_{\rm cl})$], and the size of
cells in this grid. 

We have shown that should the detection limit be underestimated,
artificially shortened {\dts}s would result. This is so because there
is a degeneracy between incompleteness and secular evolution, i.e.,
the fading-driven turn-over in a CMF in a given age bin is interpreted as
resulting from secular evolution, leading to a shortened cluster
disruption time-scale.

Having set the best possible constraints on $t_4^{\rm dis}$, we
explored the corresponding CFR, in particular considering $t_4^{\rm
dis}=1$ and 10 Gyr.  The CFR has been increasing steadily from about 0.3
clusters Myr$^{-1}$ 5 Gyr ago, to a present rate of $(20-30)$ clusters
Myr$^{-1}$, for clusters spanning an initial mass range of $\sim 100-10^7$ M$_\odot$. The
CFRs inferred for both disruption time-scales differ by at most a
factor of three (Fig.~\ref{fig:ad_all}). At older ages however, the
situation becomes unclear. The uncertainty in the CFR as a result of
the uncertainty on $t_4^{\rm dis}$ increases. In addition, the overall
temporal behaviour of the CFR depends on the shape of the ICMF of the
oldest, globular cluster-like objects. If this is the universal
Gaussian ICMF, then the CFR has increased steadily over a Hubble time
from $\sim 1$ cluster Gyr$^{-1}$ 15 Gyr ago to its present value. On
the contrary, if the ICMF has always been a power law with a slope
close to $\alpha=-2$, the CFR exhibits a minimum some 5 Gyr ago. Our
results may be related to the orbital history and dynamics of the LMC
with respect to both the SMC and the Milky Way, although this remains
poorly constrained because of a lack of proper motion measurements
with the required accuracy (Besla et al.~2007).

Additionally, we note that interactions between the Clouds and between
the Clouds and the Galaxy, while affecting their star-formation
history, also affect the {\it structure} of the Magellanic Clouds
(Bekki \& Chiba 2005). This may have induced temporal variations in
the {\dts} over the past Hubble time, rendering the {\dts} estimate at
old age even more uncertain.

Finally, we have investigated which strategy should be adopted in the
future in order to better constrain the {\dts} in the LMC.
Specifically, we have generated synthetic cluster populations defined
by a given cluster formation history, ICMF and various dissolution
rates.  After the addition of Gaussian noise to mimic an observational
situation, we processed these simulated data sets in the same way as
the actual LMC data. We demonstrate that {\it if} the {\dts} is known,
then the CFR can be derived accurately. We confirm our inability to
distinguish $t_4^{\rm dis}=1$ Gyr from 10 Gyr because of the bright
detection limit. With such a bright detection limit, the expected
turn-over in the CMF caused by dynamical evolution is not detected,
for any cluster age range. As a result, only a lower limit to the
{\dts} can be retrieved, i.e., we can exclude all $t_4^{\rm dis}$ that
are short enough to lead to a turn-over above the detection limit.
To obtain age and mass estimates for an LMC star cluster sample complete
above $M_V^{\rm lim}=-3.5$ is desirable to more easily distinguish between
$t_4^{\rm dis}=1$ Gyr and $t_4^{\rm dis}=10$ Gyr.

\section*{Acknowledgments}
We thank Peter Anders for helpful discussions and for performing 
the broad-band SED analysis
resulting in the cluster age and mass determinations used in this
paper. GP acknowledges support from the Belgian Science Policy Office
in the form of a Return Grant and from the Alexander von Humboldt
Foundation in the form of a Research Fellowship. We are grateful for
research support and hospitality at the International Space Science
Institute in Bern (Switzerland), as part of an International Team
Programme. We acknowledge partial financial support from the UK's
Royal Society through an International Joint Project grant aimed at
facilitating networking activities between the universities of
Sheffield and Bonn. This research has made use of NASA's Astrophysics
Data System Abstract Service.

\label{lastpage}

\end{document}